\documentclass[useAMS,usenatbib]{mn2e}

\usepackage{graphicx,subfigure} % subfigure is needed to put multiple plots in one figure
\usepackage{amsmath}

\newcommand\aj{\ref@jnl{AJ}}%
\newcommand\actaa{\ref@jnl{Acta Astron.}}%
\newcommand\araa{\ref@jnl{ARA\&A}}%
\newcommand\apj{\ref@jnl{ApJ}}%
\newcommand\apjl{\ref@jnl{ApJ}}%
\newcommand\apjs{\ref@jnl{ApJS}}%
\newcommand\ao{\ref@jnl{Appl.~Opt.}}%
\newcommand\apss{\ref@jnl{Ap\&SS}}%
\newcommand\aap{\ref@jnl{A\&A}}%
\newcommand\aapr{\ref@jnl{A\&A~Rev.}}%
\newcommand\aaps{\ref@jnl{A\&AS}}%
\newcommand\azh{\ref@jnl{AZh}}%
\newcommand\baas{\ref@jnl{BAAS}}%
\newcommand\caa{\ref@jnl{Chinese Astron. Astrophys.}}%
\newcommand\cjaa{\ref@jnl{Chinese J. Astron. Astrophys.}}%
\newcommand\icarus{\ref@jnl{Icarus}}%
\newcommand\jcap{\ref@jnl{J. Cosmology Astropart. Phys.}}%
\newcommand\jrasc{\ref@jnl{JRASC}}%
\newcommand\memras{\ref@jnl{MmRAS}}%
\newcommand\mnras{\ref@jnl{MNRAS}}%
\newcommand\na{\ref@jnl{New A}}%
\newcommand\nar{\ref@jnl{New A Rev.}}%
\newcommand\pra{\ref@jnl{Phys.~Rev.~A}}%
\newcommand\prb{\ref@jnl{Phys.~Rev.~B}}%
\newcommand\prc{\ref@jnl{Phys.~Rev.~C}}%
\newcommand\prd{\ref@jnl{Phys.~Rev.~D}}%
\newcommand\pre{\ref@jnl{Phys.~Rev.~E}}%
\newcommand\prl{\ref@jnl{Phys.~Rev.~Lett.}}%
\newcommand\pasa{\ref@jnl{PASA}}%
\newcommand\pasp{\ref@jnl{PASP}}%
\newcommand\pasj{\ref@jnl{PASJ}}%
\newcommand\qjras{\ref@jnl{QJRAS}}%
\newcommand\rmxaa{\ref@jnl{Rev. Mexicana Astron. Astrofis.}}%
\newcommand\skytel{\ref@jnl{S\&T}}%
\newcommand\solphys{\ref@jnl{Sol.~Phys.}}%
\newcommand\sovast{\ref@jnl{Soviet~Ast.}}%
\newcommand\ssr{\ref@jnl{Space~Sci.~Rev.}}%
\newcommand\zap{\ref@jnl{ZAp}}%
\newcommand\nat{\ref@jnl{Nature}}%
\newcommand\iaucirc{\ref@jnl{IAU~Circ.}}%
\newcommand\aplett{\ref@jnl{Astrophys.~Lett.}}%
\newcommand\apspr{\ref@jnl{Astrophys.~Space~Phys.~Res.}}%
\newcommand\bain{\ref@jnl{Bull.~Astron.~Inst.~Netherlands}}%
\newcommand\fcp{\ref@jnl{Fund.~Cosmic~Phys.}}%
\newcommand\gca{\ref@jnl{Geochim.~Cosmochim.~Acta}}%
\newcommand\grl{\ref@jnl{Geophys.~Res.~Lett.}}%
\newcommand\jcp{\ref@jnl{J.~Chem.~Phys.}}%
\newcommand\jgr{\ref@jnl{J.~Geophys.~Res.}}%
\newcommand\jqsrt{\ref@jnl{J.~Quant.~Spec.~Radiat.~Transf.}}%
\newcommand\memsai{\ref@jnl{Mem.~Soc.~Astron.~Italiana}}%
\newcommand\nphysa{\ref@jnl{Nucl.~Phys.~A}}%
\newcommand\physrep{\ref@jnl{Phys.~Rep.}}%
\newcommand\physscr{\ref@jnl{Phys.~Scr}}%
\newcommand\planss{\ref@jnl{Planet.~Space~Sci.}}%
\newcommand\procspie{\ref@jnl{Proc.~SPIE}}%
\newcommand\lsim{\la}
\newcommand\gsim{\ga}

\newcommand \Chandra{{\sl Chandra}}

\newcommand \mum{ \mu{\rm m} }
\newcommand \charsig{ \overset{\sim}{\sigma} }

\newcommand \amin{ a_{\rm min} }
\newcommand \amax{ a_{\rm max} }

\newcommand \NH{{\rm N}_{\rm H}}
\newcommand \NHone{{\rm N}_{\rm H,1}}
\newcommand \NHtwo{{\rm N}_{\rm H,2}}
\newcommand \col{{\rm cm}^{-2}}

\newcommand \tsca{\tau_{\rm sca}}
\newcommand \Fps{F_{ps}}
\newcommand \Fh{F_{h}}
\newcommand \alphmin{\alpha_{1}}
\newcommand \alphmax{\alpha_{2}}

\begin{document}

\title[The dust scattering halo of Cygnus X-3]{The dust scattering halo of Cygnus X-3}
\author[Corrales \& Paerels]
{
L.~R.~Corrales$^{1,2}$, F.~Paerels$^{3}$ \\
$^1$MIT Kavli Institute for Astrophysics and Space Research\\
$^2$Columbia University\\
$^3$Columbia Astrophysics Laboratory
}

\maketitle
\begin{abstract}

Dust grains scatter X-ray light through small angles, producing a diffuse halo image around bright X-ray point sources situated behind a large amount of interstellar material.  We present analytic solutions to the integral for the dust scattering intensity, which allow for a Bayesian analysis of the scattering halo around Cygnus X-3.  
Fitting the optically thin 4-6~keV halo surface brightness profile yields the dust grain size and spatial distribution.  We assume a power law distribution of grain sizes ($n \propto a^{-p}$) and fit for $p$, the grain radius cut-off $\amax$, and dust mass column.  We find that a $p \approx 3.5$ dust grain size distribution with $\amax \approx 0.2~\mum$ fits the halo profile relatively well, whether the dust is distributed uniformly along the line of sight or in clumps.  We find that a model consisting of two dust screens, representative of foreground spiral arms, requires the foreground Perseus arm to contain 80\% of the total dust mass.  The remaining 20\% of the dust, which may be associated with the outer spiral arm of the Milky Way, is located within 1~kpc of Cyg X-3.  
Regardless of which model was used, we found $\tau_{\rm sca} \sim 2 \ E_{\rm keV}^{-2}$.  
We examine the energy resolved halos of Cyg X-3 from 1 - 6~keV and find that there is a sharp drop in scattering halo intensity when $E < 2-3$~keV, which cannot be explained with multiple scattering effects.  We hypothesize that this may be caused by large dust grains or material with unique dielectric properties, causing the scattering cross-section to depart from the Rayleigh-Gans approximation that is used most often in X-ray scattering studies.  The foreground Cyg OB2 association, which contains several evolved stars with large extinction values, is a likely culprit for grains of unique size or composition.

\end{abstract}


%---------------------------------------------------
\section{Introduction}
\label{sec:Introduction}

Dust grains are a vital part of the interstellar medium (ISM), aiding in gas cooling for star formation, providing a site for chemical reactions, and acting as the seeds for planetesimal growth.  ISM dust is typically observed in absorption, over the UV and optical, or emission in the infrared.  However, high energy studies of interstellar dust grains complement information at other wavelengths for several reasons.  First, the dust scattering cross-section in the X-ray is highly sensitive to grain radius ($a$), making it ideal for gauging the large end of the size distribution.  This is more difficult to do at other wavelengths.  Large grains $\sim 0.3~\mum$ in radius have a flat extinction efficiency for UV and optical light, which affects the normalization of the extinction curve and not its slope.  Infrared emission ($\lambda \lsim 50\ \mum$) also tends to be dominated by the smallest carbonaceous grains (PAHs); the larger $0.1 \ \mum$ grains will glow at these wavelengths only when subjected to intense radiation \citep{Draine2007}.  Second, dust grains are relatively transparent to X-rays \citep{Wilms2000}.  As a consequence, X-rays probe the full abundance of interstellar elements (gas plus dust) and higher ISM column densities  than UV or optical studies.

We focus here on dust scattering of X-rays over small angles, which produces a diffuse halo image around bright X-ray point sources \citep{Over1965}.  The first X-ray scattering halo was imaged with the Einstein Observatory around 4U$1658-48$ \citep{Rolf1983}.  Since then, scattering halos have been observed around various Galactic X-ray binaries \citep[e.g.][]{Witt2001,Smith2008}, anomalous X-ray pulsars \citep{Tiengo2010}, and gamma-ray bursts that pass through dusty regions of the Milky Way \citep{Vau2006}.

It has been shown that a power law distribution ($N_d \propto a^{-p}$ with $p \approx 3.5$) with a mix of graphite and silicate grains reproduce extinction curves in the UV and optical \citep[][hereafter MRN]{MRN1977}.  
Updated grain size distributions beyond the simple power law regime have been developed to better reproduce infrared and microwave emission features.  
\citet[][hereafter WD01]{WD2001} produced a distribution that mainly modified the lower end of the grain size distribution, which shines brightest in the infrared. 
\citet[][hereafter ZDA]{Zubko2004} did the same, but aimed to preserve elemental abundance constraints.  They developed a series of models with varying mixtures of silicate, graphite, amorphous carbon, PAH, and composite material sometimes referred to as ``fluffy'' dust.
Several authors have shown that ZDA models fit X-ray scattering halos comparatively well to slightly better than MRN and WD01 size distributions \citep{Smith2006, Smith2008, Valencic2009}.  However, the best fitting ZDA models tended to be BARE-GR and BARE-AC (solid grains of graphite and amorphous carbon), which are more similar to MRN than the ZDA COMP (fluffy grain) models.  
The goal of this work is to examine how X-ray scattering can help drive our understanding of dust grain size distribution.  We take power law models as a jumping off point, and ask the question: what is the maximum grain size cut-off that can explain X-ray scattering from the diffuse ISM?

We examine one of the brightest X-ray scattering halos available in the {\sl Chandra} archive, associated with the high mass X-ray binary (HMXB) Cygnus X-3.  We describe the dust scattering physics and foreground ISM environment below.  The observation and PSF subtraction method is described in Section~\ref{sec:Observation}.  We take the Bayesian approach to fitting the dust grain size distribution in Section~\ref{sec:HaloFit}, using two models for the spatial distribution of dust along the line of sight.  In one case we assume dust is uniformly distributed; in the other we model the halo with two infinitesimally thin dust screens.  In Section~\ref{sec:HaloRatio} we present flux measurements and model residuals from energy resolved scattering halos between 1 and 6~keV.  For completeness, we report in Section~\ref{sec:Alternative} two alternate fits to the 4-6~keV dust scattering halos and the model implications for optical and UV extinction. In Section~\ref{sec:Comparison}, we compare our results to other papers that study ISM on the Cyg X-3 sight line.  Conclusions are summarized in Section~\ref{sec:Conclusion}.  Finally, the Appendix contains the analytic solution for the halo intensity in the case of a power law distribution of dust grain sizes.  These solutions allow for fast computation of the halo surface brightness profile, making Bayesian analysis feasible.

\begin{figure}
\begin{center}
	\includegraphics[scale=0.3, trim=0 0 0 0]{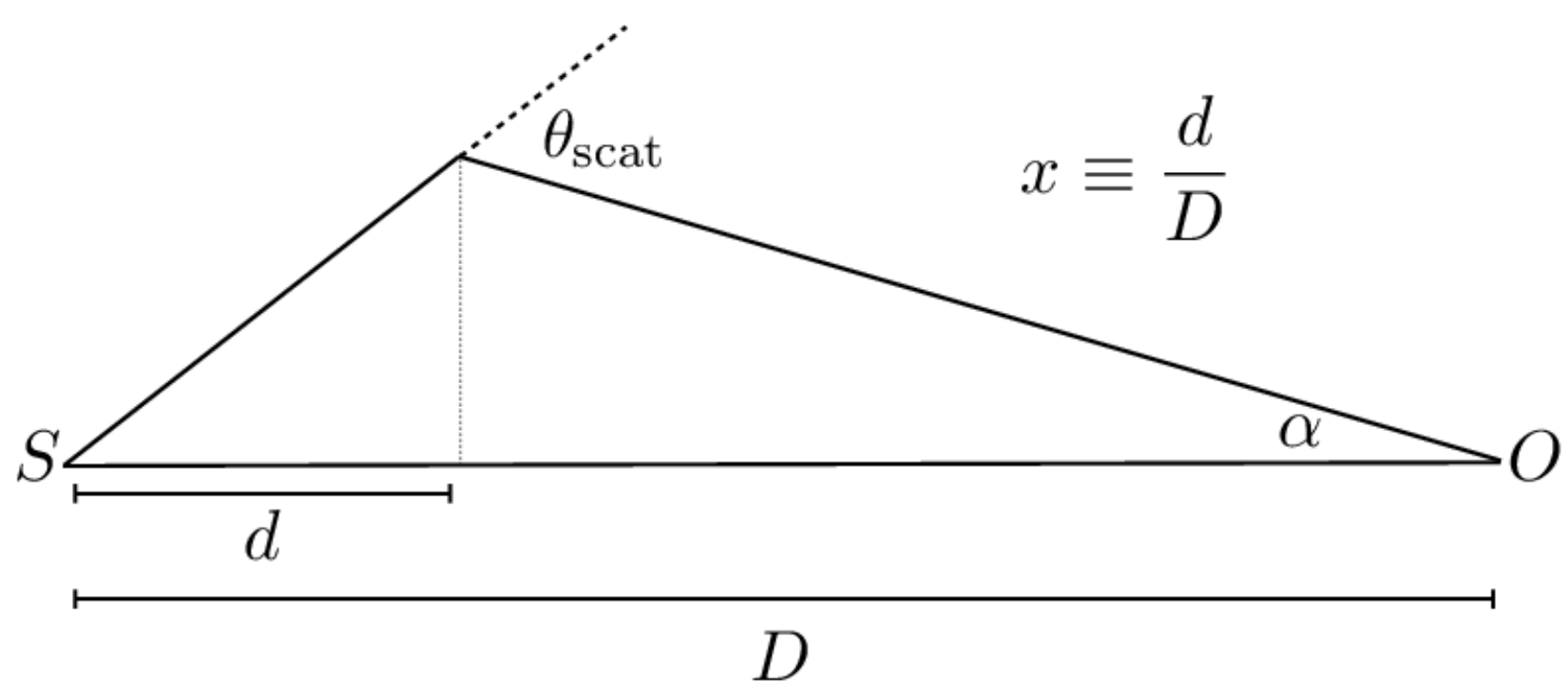}
	\caption{ Geometry of X-ray scattering through the interstellar medium, with an X-ray point source at $S$ and observer at $O$. }
	\label{fig:SOdiagram}
\end{center}
\end{figure}

%----
\subsection{Dust scattering physics}
\label{sec:DustScattering}

Because Cyg X-3 is situated behind a very large column of gas and dust, it is dark for $E < 1$ keV.  Assuming MRN-type dust, our calculations for the dust scattering cross-section are safely in the Rayleigh-Gans (RG) regime of $E_{\rm keV} \gsim a_{\mum}$, where $a$ is the dust grain radius in units of $\mum$ \citep{SD1998}.%\footnote{We find that this is true when using an MRN-like power law distribution of dust grains with $\amax = 0.25\ \mum$.  As shown by \citet{SD1998}, the RG approximation departs from Mie around 2 keV with a single size $0.25 \ \mum$ grain distribution.}  
We follow \citet{SD1998} in applying the Drude approximation for the complex index of refraction, which treats each dust grain as a sphere of free electrons.  Assuming that there are on average two baryons in the atomic nucleus for every electron,  the total dust scattering cross-section is
\begin{equation}
\label{eq:red_Sigma}
	\sigma_{\rm sca} \approx 6.2 \times 10^{-7} \  
	\rho_3^2 \ a_{\mum}^4 \ E_{\rm keV}^{-2} \ {\rm cm}^2
\end{equation}
where $\rho_3 = \rho / (3\ {\rm g\ cm}^{-3})$, $a_{\mum} = a / (1\ \mum)$, and $E_{\rm keV} = E / (1 \ {\rm keV})$ are typical values.

The RG differential scattering cross-section contains a first order Bessel function that can be approximated with a Gaussian \citep{MG1986}.  Using the Drude approximation again, 
\begin{equation}
\label{eq:red_dSigma}
	\frac{d\sigma_{\rm sca}}{d\Omega} \approx 1.13 \ 
	\exp \left( \frac{ -\theta_{\rm sca}^2 }{ 2 \charsig^2 } \right) \ 
	\rho_3^2 \ a_{\mum}^6 \ {\rm cm}^2 \ {\rm ster}^{-1}
\end{equation}
with the characteristic width
\begin{equation}
\label{eq:charsig}
	\charsig = \frac{ 1.04\ {\rm arcmin} }{ E_{\rm keV}\ a_{\mum} }
\end{equation}

The intensity of the dust scattering halo is calculated by integrating the scattering cross-section along the line of sight (Figure~\ref{fig:SOdiagram}).  A patch of dust grains at $d \equiv x D$ distance away from the X-ray source at $S$,\footnote{Note that in other publications involving X-ray scattering, $x$ is often used to denote the distance between the {\sl observer} ($O$) and the dust grains.  In that case, all instances of $x$ should be replaced by $(1-x)$ throughout.  We chose to define $x$ as shown in Figure~\ref{fig:SOdiagram} for mathematical elegance.} will see a flux $L_S / 4\pi d^2 = F_a / x^2$ where $F_a$ is the apparent flux at point $O$.  Light observed at angle $\alpha$ requires $\theta_{\rm sca} = \alpha / x$.  
Using $N_d$  for the dust column density and $\xi(x)$ to parameterize the density as a function of position along the line of sight, we get
\begin{equation}
\label{eq:halo_intensity}
	I_h(\alpha,E) = \int_a \int_0^1 \frac{ F_a }{ x^2 } 
		\frac{d\sigma}{d\Omega} \left( \theta_{\rm sca} = \frac{\alpha}{x} \right)
		N_d \xi(x) dx da.
\end{equation}
The dust scattering code used in this work calculates the scattering halo as normalized by apparent source flux,
\begin{equation}
\label{eq:halo_par}
	\frac{d\psi_h}{d\Omega} \left(\alpha, E\right) = \frac{I_h(\alpha, E)}{F_a}
\end{equation}
which will be used later in Section~\ref{sec:HaloRatio}.

The effect of absorption should be considered carefully because the scattered light takes a longer path \citep[e.g.][]{TS1973}:
\begin{equation}
\label{eq:pathdiff}
	\delta x = \frac{\alpha^2 (1-x)}{2 x}.
\end{equation}
Assuming that the ISM is homogeneous enough that the scattered light path does not differ significantly in extinction properties, dust scattered light is subject to an additional $\delta \tau_{\rm abs} \approx \tau_{\rm abs} \delta x$, where $\tau_{\rm abs}$ is the total absorption along distance $D$.
By the nature of small angle scattering, the observer at $O$ will mostly view dust that is at intermediate distances, $x \sim 1/2$, making $\delta x \approx \alpha^2/2$.
The largest observation angles in this study are $\alpha \sim 100''$, resulting in $\delta x \lsim 10^{-7 }$.  This is negligible even for ISM columns where $N_{\rm H} \gsim 10^{22}$ cm$^{-2}$, which have $\tau_{\rm abs} \gsim 1$\citep{Wilms2000}.
Thus from here on forward we combine the absorption term with $F_a$ so that it represents the absorbed apparent flux as it is incident on the observer: $F_a \equiv L_s\ e^{-\tau_{\rm abs}} / 4 \pi D^2$.

We consider two cases: a uniform distribution of dust grains along the line of sight, $\xi (x) = 1$, 
and an infinitesimally thin screen of dust grains at position $x_s$, so that $\xi (x) = \delta(x-x_s)$.

%----
\subsection{ISM column}
\label{sec:ISMcolumn}

Cyg X-3 is located in the galactic plane at $(l,b) = (79.8, +00.7)$.  At a distance of 7-13 kpc from the Sun \citep{Dickey1983,Predehl2000}, the HMXB is located %near the edge of the Galactic disk, 
behind one or two spiral arms of the Milky Way --  Perseus and the outer arm \citep{Russeil2003, Reid2014}.  
Cyg X-3 is also situated behind the young stellar association Cyg OB2, which is associated with the larger Cygnus X molecular region \citep{Dame2001, Knodlseder2003, Wright2015}.  
This particular sight line thereby offers a unique laboratory for probing dust physics within different ISM phases using the phenomenon of X-ray scattering.

Radio surveys of the 21-cm line give a neutral hydrogen column towards Cyg X-3 of $N_{\rm HI} \gsim 10^{22}$ cm$^{-2}$ \citep{LABsurvey}.  The Milky Way CO survey shows that the Cyg X-3 sightline is particularly patchy, but contains a total proton density $N_{\rm H} \approx 10^{22}$ cm$^{-2}$ in the form of molecular hydrogen \citep{Dame2001}.  Taking both these measurements as a lower limit, because Cyg X-3 seems to be near the edge of the Galaxy and because we expect some portion of the ISM hydrogen to be ionized, the total ISM column is likely $N_{\rm H} > 2 \times 10^{22}$ cm$^{-2}$.

X-ray spectroscopy offers another means to measure the ISM column density.  Absorption by neutral hydrogen via the photoelectric effect dominates below 1 keV, but the metal content of the ISM accounts for a considerable fraction of the total absorption above 1 keV \citep{Wilms2000}.  
\citet{PS1995} used ROSAT observations of Cyg X-3 to estimate $N_{\rm H} \approx 3-4 \times 10^{22}$ cm$^{-2}$, which is consistent with the above radio surveys.  They also estimate the total optical depth to scattering for Cyg X-3 is $\tsca \sim 1.5~E_{\rm keV}^{-2}$, which means that the extinctive effects of dust scattering must be incorporated into spectral models for ISM extinction.  This is particularly relevant for the high resolution optics of \Chandra~because scattering removes light from the source extraction region. 

%---------------------------------------------------
\section{Data Reduction}
\label{sec:Observation}

 The $0.5''$ per pixel resolution and low background makes {\sl Chandra} the best X-ray observatory available for imaging dust scattering halos \citep{Chandra}.    We chose to analyze the longest {\sl Chandra} observation (50 ks) of Cygnus X-3, ObsId 6601, which was taken with the High Energy Transmission Grating \citep[HETG,][]{HETG}.  The data presented in this work was extracted using common data reduction procedures and CIAO version 4.5.

%----
\subsection{Fit to the HETG Spectrum}
\label{sec:SpectralFit}

%---------------------------------------------------
\begin{figure}
\begin{center}
	\includegraphics[scale=0.62, trim=0 0 0 0]{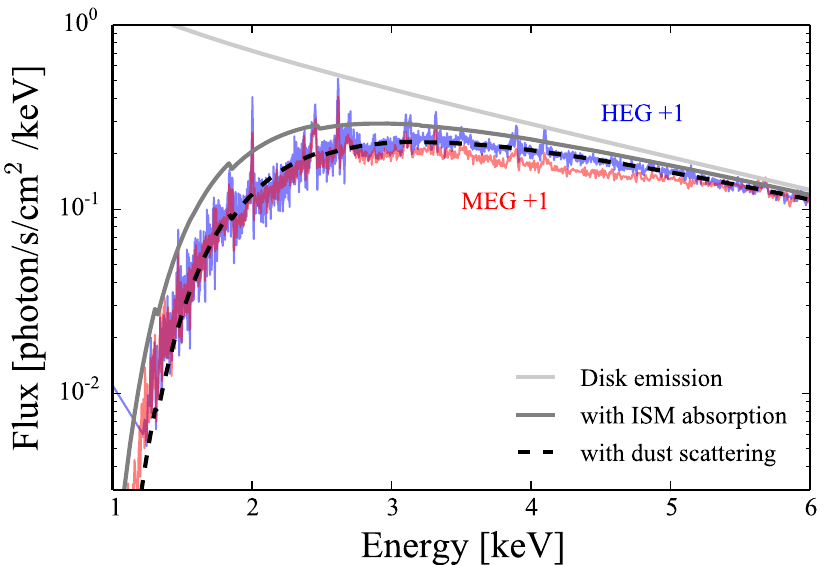}
	\caption{ The 1-6~keV continuum fit to the high resolution spectrum of Cygnus X-3 uses a \texttt{diskpn} model for the accretion disk emission, \texttt{TBnew} for the effect of ISM absorption, and a dust extinction model utilizing RG-Drude scattering. }
	\label{fig:CygX3Spectrum}
\end{center}
\end{figure}
%---------------------------------------------------

We used the \texttt{tg\_findzo} tool in CIAO to center the HETG extraction regions on the point source, then used standard calibration methods to get a spectrum of Cyg X-3, shown in Figure~\ref{fig:CygX3Spectrum}.  Some disagreement between the MEG and HEG spectrum in the 3-5~keV energy range is an indication of pileup (discussed in more detail in Section~\ref{sec:PSFFit}), because light is dispersed over smaller angles with the MEG as compared to the HEG.  To test for possible non-linear effects in the HEG arms, we applied the pileup correction \texttt{simple\_gpile2} \citep{simple_gpile,simple_gpile2} with the physical spectral models described below.  We found that the MEG and HEG arms could be fit simultaneously with less than 1\% of HEG counts being affected by pileup.  Pileup is complicated by the fact that Cyg X-3 is a variable source (\S\ref{sec:variability}).  Based on the raw counts taken from time intervals when Cyg X-3 is at its brightest, which is about 20\% of the full exposure time, the HEG spectrum will experience $\leq 7\%$ pileup in the region around 4-5~keV. From here forward we choose to model Cyg X-3 using the HEG arms only, which we presume is accurate on the 5\% level.

The magnitude of ISM extinction prohibits precise modeling of the the unabsorbed spectrum of Cyg X-3 over the {\sl Chandra} energy range.  We found that a pseudo-Newtonian accretion disk emission model appropriate for many HMXBs, \texttt{diskpn} \citep{Gierl1999}, fit the spectrum slightly better than a single power law.  However, we did not attempt to model other effects such as Compton reflection or hybrid plasmas, which would describe the $E > 6$~keV component of the spectrum \citep[e.g.][]{SZM2008, Zdz2010}.  The continuum extinction model consists of two components: neutral ISM absorption \citep[\texttt{TBnew},][]{Wilms2000, Juett2004, Juett2006} and dust scattering, which will remove light from the spectrum extraction region.  As a first order approximation, we use a custom dust scattering model that follows $\exp(-\tsca)$, where $\tsca$ was calculated by choosing a dust-to-gas mass ratio typical of the Milky Way \citep[0.009, e.g.][]{DraineBook} and tying it to the ISM column in the following way:
\begin{equation}
\label{eq:DustExtinctionComponent}
	\tau_{\rm sca} \approx 0.009\ N_{\rm H}\ m_p\ \kappa_{\rm keV}\ E_{\rm keV}^{-2}.
\end{equation}
Here, $\kappa_{\rm keV} \approx 3.3 \times 10^{3}$~g$^{-1}$~cm$^2$ is the 1~keV scattering opacity from an MRN distribution of dust with $\amax = 0.25~\mum$, and $N_{\rm H}$ is the ISM column from \texttt{TBnew}.

Figure~\ref{fig:CygX3Spectrum} shows the relative contributions of the source model, ISM absorption, and dust scattering for a fit to the 1-6~keV energy range.  The best fit ISM column is $N_{\rm H} \approx 4.3  \times 10^{22}$~cm$^{-2}$, implying $\tau_{\rm sca} (1~{\rm keV}) \approx 2$.  This is roughly consistent with the results of the \citet{PS1995} study.

We also noted through the course of examination that the fit to the broad 1-6~keV energy band shown in Figure~\ref{fig:CygX3Spectrum} is systematically low by about 10\% around 2~keV and 40\% at the softest energy, around 1.25 keV.   This is due to the time variation in Cyg X-3, which experiences absorption from its stellar companion (\S\ref{sec:variability}), so the time integrated spectrum may be better represented by a partial covering model.  For the purposes of creating a PSF template and extracting a scattering halo, we simply need an accurate measurement of the photon flux.  We fit a power law to each 0.5~keV wide energy band between 1 and 3~keV to to get the flux for these respective bins.  
%These piece-wise power law fits shifted the measured flux between 10\% (around 2~keV) and 40\% (around 1.25~keV) with respect to the fit shown in Figure~\ref{fig:CygX3Spectrum}.

%----
\subsection{PSF templates}
\label{sec:PSFFit}

The High Resolution Mirror Assembly (HRMA) on {\sl Chandra} focuses about 90\% of the X-ray light into a few pixel ($2''$) region.  The point spread function (PSF) is composed of two parts: the core, where the majority of light is focused, and the wings, where light is spread diffusely due to scattering off of fine surface features in the mirror.  Correctly subtracting the PSF from the image is of utmost importance for determining scattering halo profile, the brightness of which is often on the order of the PSF wing brightness.  The PSF can be simulated with {\sl Chandra} calibration tools (ChaRT and MARX), but these methods are known to under-predict the wing brightness by a factor of 2-10 \citep{Smith2002}.

A PSF template may be created from a bright source with a relatively dust-free sight line, but the situation is complicated by the non-linear response of the CCD detectors on {\sl Chandra}.  When more than one X-ray photon hits a pixel before the CCD is read out, the resulting electron cloud will be interpreted as either a single photon of larger energy or as a cosmic ray.  This effect is known as pileup, and it occurs for any source with a flux on the order of one photon per readout time (typically 3.2 seconds).  For grating-free observations, pileup prevents the correct normalization of a PSF template.  However, pileup is mitigated for sources imaged with the HETG in place, both because the effective area is reduced and because a more accurate pile-up free spectrum can be extracted from the grating-dispersed light.

We chose an HETG observation of 3C 273 (ObsId 459) to create a template of the PSF wings.  3C 273 is a bright quasar situated above the Galactic plane; consequently it has a low ISM column \citep[$N_H \approx 10^{20}$ cm$^{-2}$, inferred from] []{SFD1998}.  
We built a PSF template by extracting a surface brightness profile of the zeroth order image using $\Delta E = 0.5$~keV bins between 1 and 6~keV.\footnote{
3C 273 has a small jet feature, which was masked out of the surface brightness profile.}
The average background level for each image was estimated from a region of the S3 chip, about $5.5'$ from 3C 273, then subtracted.  Using a power law fit to the HETG spectrum, we also made 0.1~keV binned weighted exposure maps for each image.  To create a template, each background subtracted surface brightness profile ($\mathcal{SB}$) was normalized by the effective area from the exposure maps, determined from a small region enclosing the point source ($\mathcal{A}_{\rm ps}$), and by the flux of 3C 273 as measured by the HETG fit ($F_{\rm ps}$):
\begin{equation}
\label{eq:TemplateEquation}
	\Psi_{\rm psf} (r, \Delta E) = \frac{ \mathcal{SB} (r, \Delta E) }{ F_{\rm ps}(\Delta E) \mathcal{A}_{\rm ps}(\Delta E) }.
\end{equation}
The PSF for Cyg X-3 was then constructed by scaling each template by the point source flux and effective area of Cyg X-3:
\begin{equation}
\label{eq:PSFEquation}
	\mathcal{SB}_{\rm psf} (r) = \sum_{E} \Psi_{\rm psf} (r, \Delta E)\ F_{\rm ps}^{X3} (\Delta E)\ \mathcal{A}^{X3}_{\rm ps} (\Delta E)
\end{equation}

%---------------------------------------------------
\begin{figure}
\begin{center}
	\includegraphics[scale=0.49, trim=30 20 0 50]{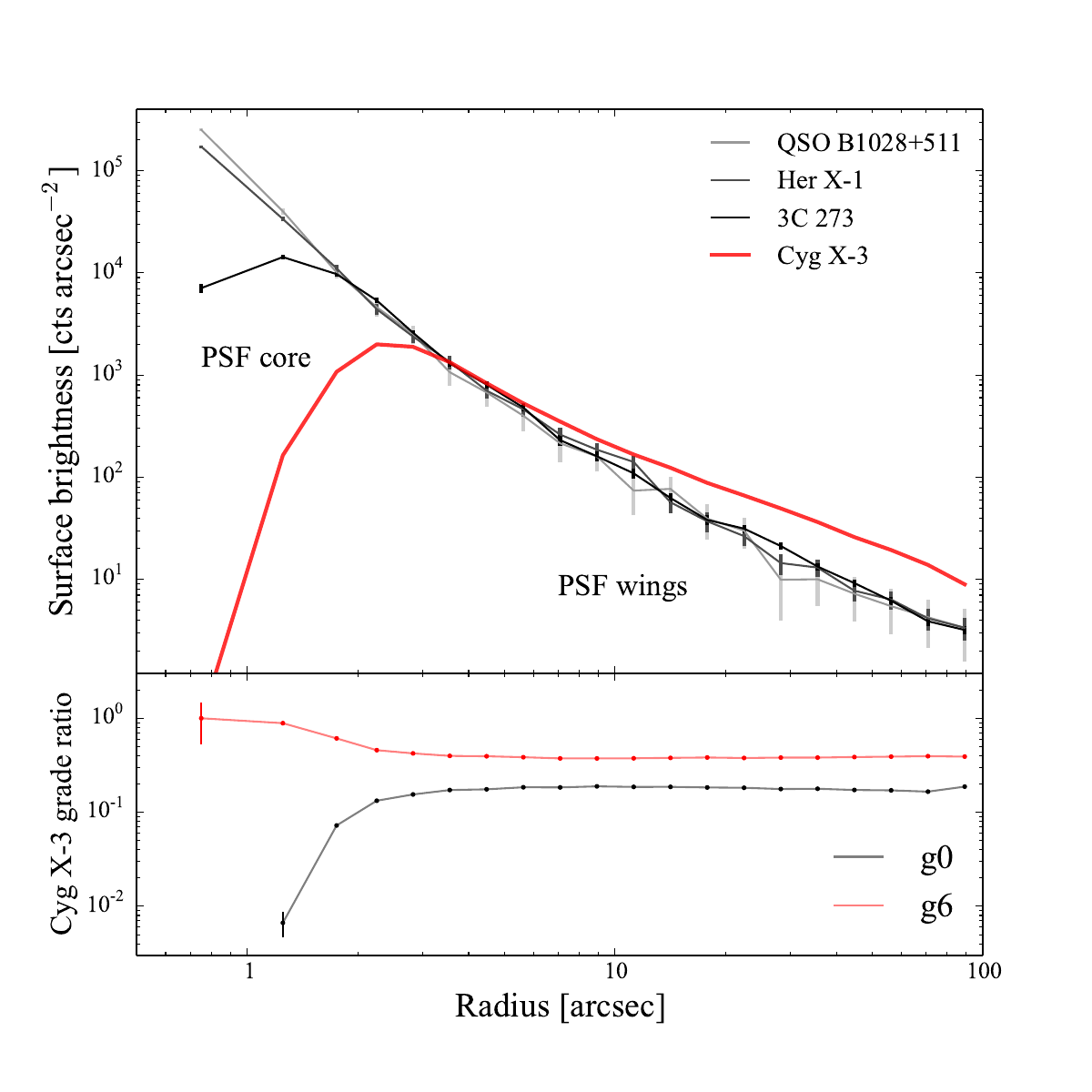}
	\caption{ 
	({\sl Top}) A comparison of PSF templates for Cyg X-3 (1-6~keV) using various dust-free sight lines. 
	({\sl Bottom}) Grade migration can be gauged by measuring the ratio of ideal (grade 0) and least acceptable (grade 6) events relative to the total number of all acceptable grades (0, 2, 3, 4, and 6).  The abrupt decrease in ideal events ({\sl black}) and increase in least acceptable events ({\sl red}) shows that pileup is mainly limited to the inner $3''$ of the Cyg X-3 surface brightness profile.
	}
	\label{fig:CompareTemplates}
\end{center}
\end{figure}
%---------------------------------------------------

As an aside, we chose to use PSF templates as opposed to the PSF models by \citet{Gaetz2004} and \citet{Gaetz2010}.  This is because, upon comparing the CCD extracted spectrum from the zeroth order image of Her X-1 (ObsId 2749) to the HETG dispersed spectrum, we found that approximately 40\% of the zeroth-order photons suffered from pileup effects.  This is much larger than the original $< 5\%$ pileup fraction surmised by \citet{Gaetz2004} based on grade migration and the number of counts per frame.  Similarly, we do not use the CXC calibration memo by \citet{Gaetz2010}\footnote{http://cxc.harvard.edu/cal/Acis/Papers/wing\_analysis\_rev1b.pdf} because we consider the spectrum determined from HETG dispersed light more reliable than that determined from the ACIS transfer streak for the purposes of normalizing the \Chandra\ PSF.  Additionally, \citet{Gaetz2010} addresses variations in the PSF wings at larger angular distances (out to $500''$) than those concerned in this work.  Our results are more sensitive to the shape of the PSF around the core-wing transition.

To demonstrate the reliability of the template method presented here, we performed the same template construction with two other dust-free sources: Her X-1 (ObsId 2749) and QSO B1028+511 (ObsId 3472).  As shown in Figure~\ref{fig:CompareTemplates}, each image suffers from varying degrees of pileup, but the PSF wings are relatively stable.  The largest deviations occur with QSO B1028+511, which is the dimmest of all three and lacks signal at higher energies.  Since 3C 273 is the brightest and has the most signal, we used it to extract the residual surface brightness profile.

%----
\subsection{Scattering halo extraction}
\label{sec:HaloExtraction}

Figure~\ref{fig:CompareTemplates} shows the surface brightness profile of Cyg X-3 relative to the PSF template.  Due to its extreme brightness, the center of Cyg X-3 is hollowed out, and much of the information is lost due to the non-linear effect of detector pileup.  We test to what extent pileup affects the wings of the Cyg X-3 profile by examining the event grades -- numbers assigned according to the shape of the electron cloud produced by a high energy event.  When multiple X-ray photons hit one region of the detector, the resultant electron cloud is likely to be asymmetric, leading to grade migration -- the assignment of a larger numeric value to what would otherwise be a normal (grade 0) event.  The bottom portion of Figure~\ref{fig:CompareTemplates} shows that grade migration is most noticeable within $3''$, which is about the point at which the surface brightness of Cyg X-3 dips below the PSF template.  \citet[][Figure~1]{McCollough} also find that the surface brightness profile of Cyg~X-3 suffers pileup fractions less than $5\%$ for $r > 3''$ and less than $1\%$ for $r > 8''$.  Pileup is therefore has a negligible effect on the dust scattering halo, which we extract for observation angles larger than $5''$.

The typical 0.5-7 keV background\footnote{{\sl Chandra} Proposer's Observing Guide, Table 6.10, http://cxc.harvard.edu/proposer/POG/} for a 50 ks observation on the S3\footnote{The ACIS S3 chip contains the zeroth order image for the nominal HETG pointing.} CCD chip is 0.02 counts pix$^{-1}$, or 0.03~counts~pix$^{-1}$ for the $5-10$~keV background.  This is two orders of magnitude below the dimmest portion of the Cyg X-3 surface brightness profile, suggesting that all the ambient light comes from X-ray scattering.  Since the quiescent detector background is minuscule by comparison, it was not included in this analysis.

There is a Bok globule located $16''$ from Cyg X-3, observed from X-ray scattering \citep{McCollough}.  A $3''$ by $4.5''$ region covering the globule contains 7120 counts, which accounts for approximately 30\% of the total brightness at that radius.  We removed the globule from the measurement, because the halo model assumes azimuthal symmetry.  We also excised from the zeroth order image a $2.5''$ wide region containing the CCD transfer streak.

The scattering halo measured from the zeroth order image might also be contaminated from the first order halo dispersed by the HETG.  To test this, we extracted a radial surface brightness profile from a rectangular region oriented in the MEG dispersion direction.  There was evidence of contamination from the MEG first order scattering halo in regions $>50''$ away from the point source center, differing by $4\sigma$ at the outermost annulus.  We therefore chose to confine the surface brightness profiles to a rectangular region perpendicular to both the MEG and HEG arms.\footnote{
Surface brightness measurements with annuli $\alpha < 20''$ receive full azimuthal coverage.}
This choice did not lead to any significant loss in signal, since the outer edges of the halo are covered by annuli of larger surface area.

%-------------------------
\begin{figure}
\begin{center}
	\includegraphics[scale=0.49, trim=30 20 0 50]{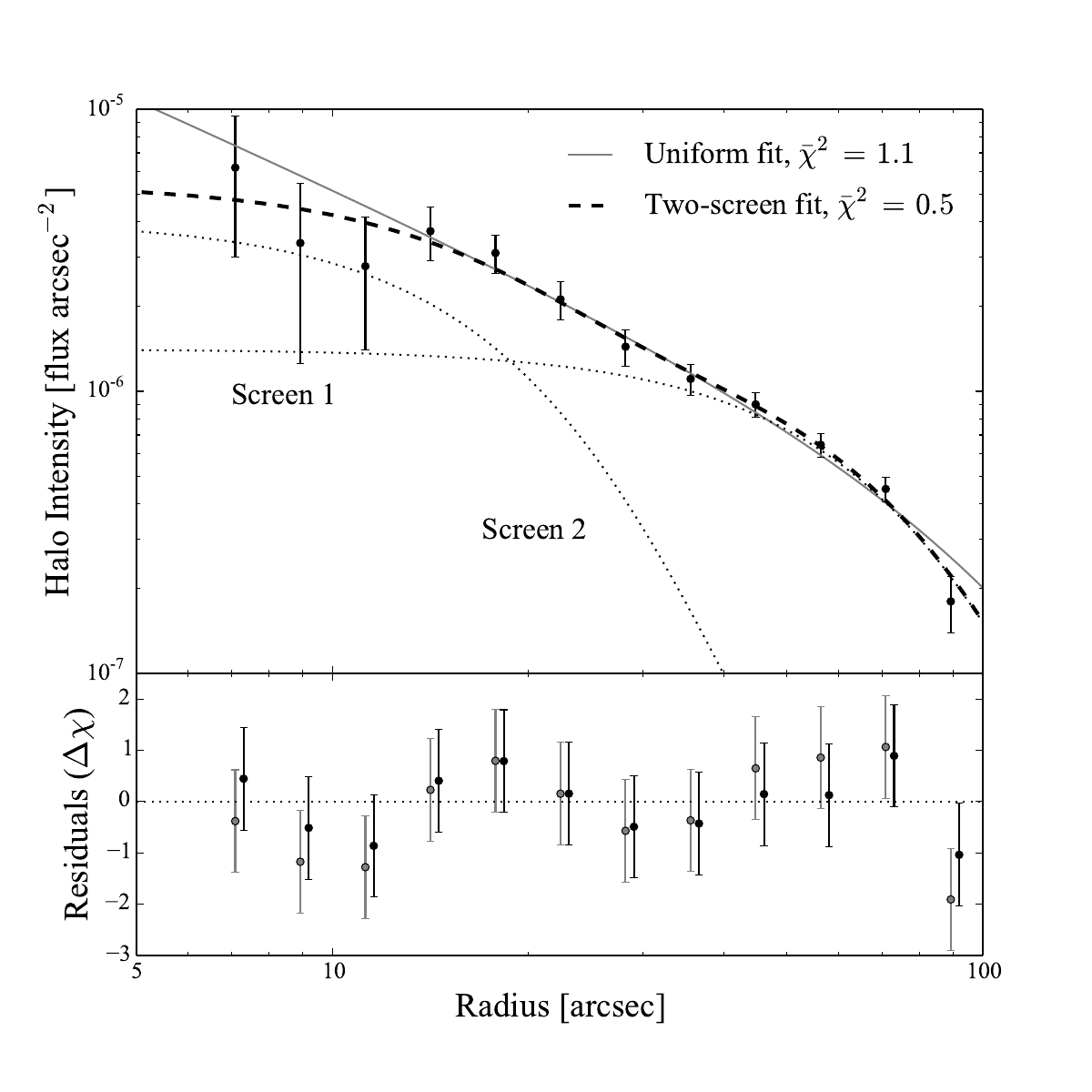}
	\caption{The 4-6~keV scattering halo intensity.  
	{\sl Top:} Overlaid are the best {\sl emcee} model fits for dust distributed uniformly along the line of sight (grey solid line) and for dust contained in two screens (black dashed line), corresponding roughly to the positions of foreground spiral arms of the Milky Way.  
	{\sl Bottom:} The residuals for the best uniform (light points) and two-screen fits (dark points) are plotted in units of sigma.  The two-screen residuals are offset horizontally to aid visibility.
	}
	\label{fig:HaloFit}
\end{center} 
\end{figure}
%-------------------------

When examining the extended image of Cyg X-3, one cannot discern whether a given photon was scattered by dust or by the {\sl Chandra} mirrors; each type is subject to a different effective area.  Aside from uneven quantum efficiencies in the CCD chip, a mirror-scattered X-ray event should be corrected by the effective area {\sl for the point source image}.  A dust-scattered X-ray event should be corrected by the effective area of the detector {\sl at the position of the event}.  Normalizing the entire image by an exposure map created with standard calibration techniques would thereby produce an inaccurate image of both the PSF and the scattering halo.  This is why we chose to scale the PSF templates by point source effective area ($\mathcal{A}_{ps}$) to obtain a raw counts surface brightness profile such as that seen in Figure~\ref{fig:CompareTemplates}.  Since all of the HETG objects used in this study used standard HETG pointings, within $30''$ of the detector focal point, we expect the overall effect of quantum inefficiencies (dead pixels and columns) to be relatively similar.  

We extracted residual surface brightness profiles, containing the raw counts from dust scattering alone, for each 0.5~keV binned image.  Then we extracted radial profiles from exposure maps using the same window described above.  The exposure maps used in this case were calculated for a single energy, not weighted, because the spectral energy distribution for the scattered light is much different from the point source.  We also could not apply weights in this case because the energy distribution should change depending on the angular distance from the point source (c.f.~Equation~\ref{eq:charsig}).  After normalizing each annulus by the mean effective area, the residual profiles were summed via
\begin{equation}
	I_h (r) = \sum_{\Delta E} \frac{ \mathcal{SB}(r, \Delta E) - \mathcal{SB}_{\rm psf}(r, \Delta E) }
		{ \mathcal{A}(r, \Delta E) }
\end{equation}
to produce the 4-6~keV halo intensity profile plotted in Figure~\ref{fig:HaloFit}.
The zeroth order image of Cyg X-3 is much more piled up than 3C 273, which does not have a hollowed out core.  We threw out data points from bins where the PSF template exceeded the observed surface brightness and kept data points with a signal-to-noise $>1$.

%%---------------------------------------------------
\section{Fit to 4-6 keV halo profile}
\label{sec:HaloFit}
%%---------------------------------------------------

Our fit to the scattering halo profile, shown in Figure~\ref{fig:HaloFit}, rests upon three fundamental assumptions.

{\sl (i) Single scattering:} The large optical depth to scattering implies that a significant fraction of photons will scatter more than once.  We take a conservative route by restricting analysis to an energy band where the scattering halo intensity is well within the optically thin, single scattering regime: $E > 4$~keV, i.e.~$\tsca \lsim 10\%$.

{\sl (ii) RG-Drude scattering from grains of a single density:} In this scattering regime, the dust grain composition is not very important because each grain is approximated as a sphere of free electrons.    We assume a grain density  $\rho = 3$ g cm$^{-3}$, which is the average between graphite and silicate materials \citep{DraineBook}.  The RG-Drude scattering cross-section is featureless and follows a power law dependence on energy.  However, the true dielectric functions will cause absorption and scattering resonances that diverge significantly from the RG-Drude approximation at low energies (\S~\ref{sec:HaloRatio}).  Restricting the energy range to 4-6~keV also alleviates the need to use the more accurate, and more computationally intensive, Mie scattering cross-section.

{\sl (iii) A power law grain size distribution:} Since a power law provides a computationally efficient means to calculate the scattering halo intensity (Equation~\ref{eq:halo_intensity}, see Appendix), we adopt the simplifying assumption that a power law is a good first order approximation to the dust grain size distribution.
For dust distributed uniformly along the line of sight ($\xi = 1$),
\begin{equation}
	I_h(\alpha, E) = \frac{ F_a }{ \sqrt{8\pi} } \ 
	\frac{ \tau_{\rm sca} }{ \alpha \charsig_0(E) } \ 
	\frac{ G_u(a,p,\alpha,E) }{ G_p(a,p) }
\end{equation}
where $\charsig_0 = 1.04'\ E_{\rm keV}^{-1}$, $G_p$ is an integral over a power law, and $G_u$ is a function of erf and incomplete gamma functions as defined in the Appendix.   For an infinitesimally thin dust screen, where $\xi = \delta(x-x_s)$,
\begin{equation}
	I_h(\alpha,E) = \frac{F_a}{x_s^2} 
		\frac{ \tau_{\rm sca} }{ 2 \pi \charsig_0^2(E) } 
		\frac{ G_s(a,p,\alpha,x_s,E) }{ G_p(a,p) }
\end{equation}
which is also described in the Appendix.  Screens produce a flat surface brightness profile, and a uniform distribution produces a cuspy profile.

As mentioned in Section~\ref{sec:SpectralFit}, the high resolution imaging capabilities of \Chandra~imply that the spectrum extracted from the HETG dispersed light is also reduced by extinction from dust scattering.  To correct for this, we modify the point source flux $\Fps$ by the model $\tsca$ value to calculate the scattering halo intensity, which is proportional to $F_a = \Fps e^{\tsca}$. % (\S\ref{sec:HaloRatio}).

We use the publicly available MCMC code {\sl emcee} to explore the parameter space in a Bayesian analysis of possible halo fits \citep{emcee}.   To see what X-ray scattering can tell us about the large end of the grain size distribution, we freeze the low end of the distribution at $0.005\ \mum$ and fit for $\amax$ and $p$.  At a fixed energy, the dust grain distribution parameters $\amax$ and $p$ mainly affect the scattering halo shape.  The optical depth to scattering mainly controls the halo normalization, which scales with the total dust mass column, $\tau_{\rm sca} = \kappa_{\rm sca}(a,p) M_d$.  We use $\NH$ as a free parameter and convert to $M_d$ using the dust-to-gas mass ratio of 0.009.%, which is the Milky Way average \citep{DraineBook}.

%------------------------------------------
\subsection{Uniform fit}
\label{sec:UniformFit}

We start with uniformly distributed dust because it has the least number of free parameters and will likely match the shape of the halo profile, which appears cuspy.  
We assigned uniform priors to $\log(\NH)$ from 14 to 24 cm$^{-2}$ and $\amax$ from $0.01 - 0.5\ \mum$.  We found that, if $p$ was allowed to take on large values, the halo fit solution became highly degenerate for $\amax > 0.3~\mum$.  A large grain size cutoff value required increasingly steep power laws,  $p > 5$, so that the small end of the size distribution greatly  dominates.  To suppress these uninformative degenerate solutions, we assigned a Gaussian prior to $p$ with mean 3.5 and standard deviation 0.5. 

%------------------------
\begin{figure*}
\begin{center}
	\includegraphics[width=\textwidth, trim=0 0 0 0]{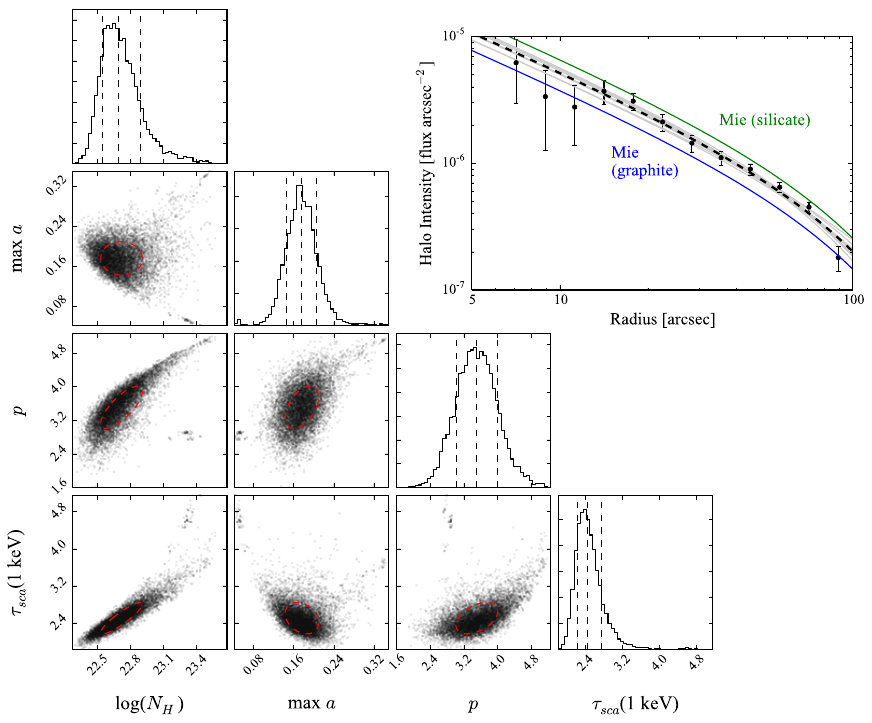}
	\caption{ The posterior distribution from {\sl emcee} is plotted with two-dimensional histograms comparing each pair of free parameters ($\NH$, $p$, and $\amax$) and the resulting $\tsca$ distribution.  There is particularly strong covariance between power-law exponent $p$ and ISM column $\NH$, making $\tsca$ covariant with both these parameters.  The vertical dashed lines in each one-dimensional histogram mark the median and 1$\sigma$ confidence interval for each parameter; the red dashed ellipses indicate the 1$\sigma$ confidence intervals in two-dimensional space.  The upper right hand figure shows the best fit {\sl emcee} walker (black dashed line) in comparison to 10 other walkers in the posterior distribution (solid grey lines).  Using the best fit walker from Table~\ref{tab:UniformFit}, the scattering halo intensities calculated with the Mie scattering cross-section for all graphite (blue) or all silicate (green) dust grains straddle the RG-Drude solution.
	}
	\label{fig:UniformTriangle}
\end{center}
\end{figure*}
%-------------------------

We used 100 walkers in {\sl emcee} to obtain $10^4$ independent samples of the posterior distribution, shown in Figure~\ref{fig:UniformTriangle}.  Two-dimensional histograms comparing each parameter against the others illustrates that they are highly covariant.  The total scattering optical depth was calculated for each sample point in the posterior distribution.
The vertical dashed lines indicate the 16th, 50th, and 84th quantiles -- corresponding to the median and 1$\sigma$ confidence interval.  Figure~\ref{fig:HaloFit} plots the best model halo corresponding to the smallest $\bar{\chi}^2$ obtained from the posterior distribution.  The best fit, median fit, and 1$\sigma$ confidence intervals are listed in Table~\ref{tab:UniformFit}.

\begin{table}
\begin{center}
\caption{Uniform fit to dust scattering halo (4-6 keV)}
\label{tab:UniformFit}
\begin{tabular}{ r c c l l  }
\hline
	& 
	{\bf Best} & 
	{\bf Median} &
	({\bf 1$\sigma$ C.I.}) & 
	{\bf Units}\\
\hline
	$\NH$ : & 3.0 & 4.9 & (3.5, 7.7) & $10^{22}$ cm$^{-2}$ \\
	$\amax$ : & 0.15 & 0.18 & (0.15, 0.21) & $\mum$ \\
	$p$ : & 2.1 & 3.5 & (3.0, 4.0) & \\
\hline
	$\tau_{\rm sca} E^2$ : & 2.2 & 2.4 & (2.2, 2.7) & keV$^{2}$ \\
\hline
\end{tabular}
\end{center}
\end{table}

%------------
\subsection{Scattering from Dust Screens}
\label{sec:ScreenFit}

\begin{table}
\begin{center}
\caption{Two screen fit to dust scattering halo (4-6 keV)}
\label{tab:ScreenFit}
\begin{tabular}{ r c c l l  }
\hline
	 &
	 {\bf Best} & 
	 {\bf Median} &
	 ({\bf 1$\sigma$ C.I.}) & 
	 {\bf Units} \\
\hline
	\multicolumn{5}{l}{Screen 1} \\
	$x_1$ : & 0.59 & 0.43 & (0.35, 0.53) &  \\
	$\NHone$ : & 1.5 & 3.3 & (2.1, 5.8) & $10^{22}$ cm$^{-2}$ \\
	\multicolumn{5}{l}{Screen 2} \\
	$x_2$ : & 0.16 & 0.12 & (0.08, 0.16) & \\
	$\NHtwo$ : & 0.3 & 0.8 & (0.4, 2.0) & $10^{22}$ cm$^{-2}$ \\
	\multicolumn{5}{l}{Dust distribution} \\
	$\amax$ : & 0.21 & 0.17 & (0.13, 0.23) & $\mum$ \\
	$p$ : & 2.4 & 3.6 & (3.1, 4.2) & \\
\hline
	Total $\NH$ : & 1.8 & 4.1 & (2.8, 6.9) & $10^{22}$ cm$^{-2}$ \\
	$\tau_{\rm sca} E^2$ : & 1.8 & 2.0 & (1.8, 2.3) & keV$^{2}$ \\
\hline
\end{tabular}
\end{center}
\end{table}

X-ray scattering can probe galactic structure in the direction of Cyg X-3, which might include features such as those associated with Cyg OB2 or Galactic spiral arms (\S\ref{sec:ISMcolumn}).  When applying the infinitesimally thin screen model, it should be kept in mind that while the total integrated halo flux will be fixed according to the optical depth of the screen, the surface brightness profile will vary according to the screen's position.  For screens closer to the observer (large $x$, see Figure~\ref{fig:SOdiagram}), the scattered flux will be spread over a large area, reducing the overall surface brightness profile.  When a screen is close to the X-ray source (small $x$), the halo surface brightness profile will be more compact and thus brighter close to the point source.

Taking 9 kpc as the best estimate for the distance to Cyg X-3, we expect the Perseus arm, about 5-6~kpc away \citep{Reid2014}, to correspond to a screen with $x_s \approx 0.4$.  The outer spiral arm, if in the foreground of Cyg X-3, will be at $x_s \leq 0.1$ and would therefore contribute most to the surface brightness profile.  The Cyg OB2 association, 1.4~kpc away \citep{Rygl2012}, corresponds to $x_s \approx 0.8$ and would contribute to the outermost portion of the profile.

We attempt to fit a model halo with the least number of dust screens, in this case two.  In the optically thin regime, the intensity of the halo from each screen can simply be added to get the total observed halo.
We ran {\sl emcee} with a six parameter model containing two screen positions ($x_1$ and $x_2$) and their respective ISM columns ($\NHone$ and $\NHtwo$), assuming the same dust grain distribution for both screens ($\amax$ and $p$).

We evaluated potential screen positions, finding that an MRN distribution of dust associated with the Cyg OB2 region would create a halo intensity $\sim 1$ count per pixel$^2$, which is too dim to contribute significantly.  We omit it from analysis.  We chose a Gaussian prior with mean 0.4 and standard deviation 0.1 applied to $x_1$ is used to model the Perseus arm.  To get at the inner portion of the scattering halo, which we believe is associated with the Milky Way outer spiral arm, we applied a Gaussian prior to $x_2$ with mean 0.05 and standard deviation 0.1.  The spread in these distributions accounts roughly for the uncertainty in the distance to Cyg X-3, $9_{-2}^{+4}$ kpc \citep{Predehl2000}.  The prior distributions on $x_1$ and $x_2$ were truncated at 0.23, so that $x_1$ always represents the screen closer to the observer.  The priors on $\NH$, $\amax$, and $p$ are identical to those in Section~\ref{sec:UniformFit}.

%------------------------
\begin{figure*}
\begin{center}
	\includegraphics[width=\textwidth, trim=0 0 0 0]{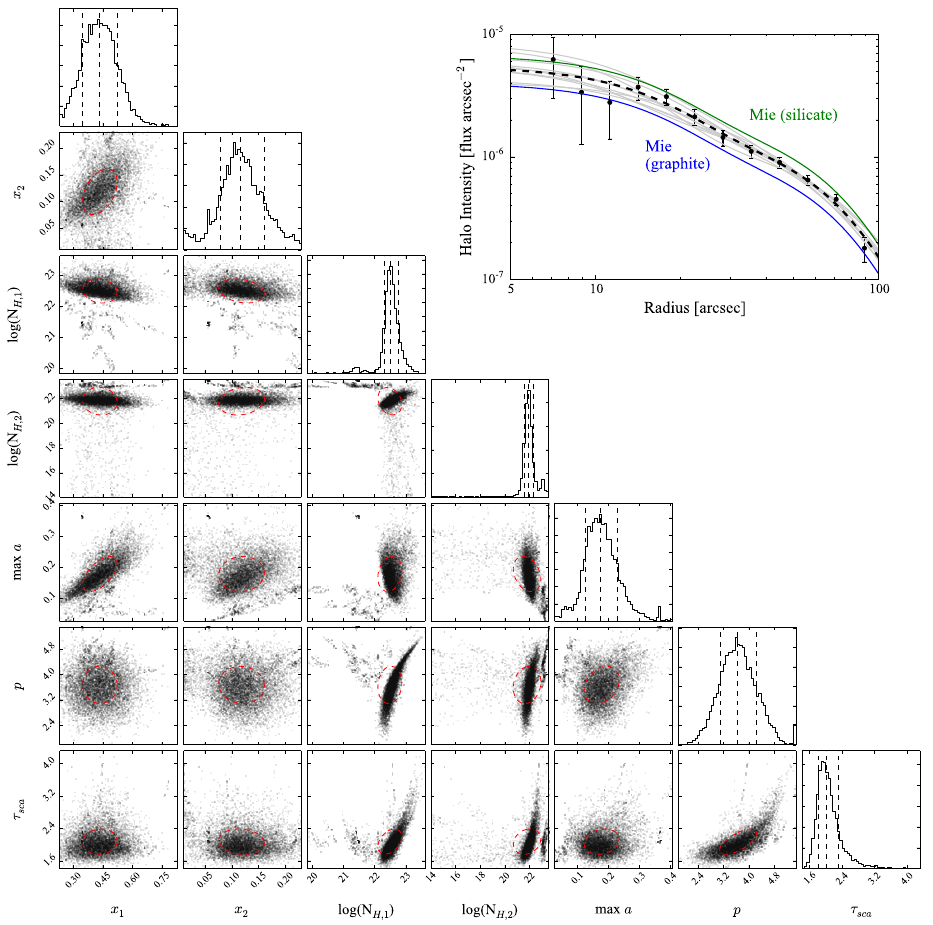}
	\caption{ The posterior distribution from {\sl emcee} is plotted with two-dimensional histograms comparing each pair of free parameters (screen positions, respective $\NH$, $p$, and $\amax$) and the resulting $\tsca$ distribution.  The vertical dashed lines in each one-dimensional histogram mark the median and 1$\sigma$ confidence interval for each parameter; the red dashed ellipses indicate the 1$\sigma$ confidence intervals in two-dimensional space.  The upper right hand figure shows the best fit {\sl emcee} walker (black dashed line) in comparison to 10 other walkers in the posterior distribution (solid grey lines).  Using the best fit walker from Table~\ref{tab:ScreenFit}, the scattering halo intensities calculated with the Mie scattering cross-section for all graphite (blue) or all silicate (green) dust grains straddle the RG-Drude solution.
	}
	\label{fig:TwoScreenTriangle}
\end{center}
\end{figure*}
%-------------------------

Figure~\ref{fig:HaloFit} shows the best fit according to smallest $\bar{\chi}^2$ drawn from the posterior distribution.  The relative contribution of each screen is also plotted.  Two-dimensional histograms displaying the covariances among each parameter are presented in Figure~\ref{fig:TwoScreenTriangle}.  Table~\ref{tab:ScreenFit} lists the best, median, and 1$\sigma$ confidence intervals for each parameter in the two screen fit, including the total dust mass and scattering optical depth.

%--------------------------------------------------------------------------
\section{Energy Resolved Scattering Halos}
\label{sec:HaloRatio}

The ratio between the scattering halo flux $\Fh = F_a (1 - e^{-\tsca})$ and the point source $\Fps = F_a e^{-\tsca}$ serves as a direct measurement for the energy dependence of the scattering cross section via 
\begin{equation}
	\label{eq:fh_fps}
	\frac{\Fh}{\Fps} = e^{\tsca} - 1.
\end{equation}
We aim to measure this quantity from the energy resolved scattering halos extracted from ObsId 6601 in the range of 1-6~keV, to test how well the 4-6~keV fits from Section~\ref{sec:HaloFit} do to approach a fully consistent halo model across a wide range of energies.

However, the field of view in ObsId 6601 is limited to about $4'$ along the axis perpendicular to the grating dispersion direction.  Our measured $\Fh/\Fps$ value will be lower because only a fraction of the total scattering halo is being captured.  Defining $\alphmin = 6.25''$ and $\alphmax = 99.5''$ from the angular limits of the extracted scattering halo profiles (Figure~\ref{fig:HaloFit}), the fraction of halo light captured is
\begin{equation}
	\label{eq:fcap}
	f_{\rm cap} (E) \equiv 
		\frac{\int_{\alphmin}^{\alphmax} I_h (\alpha, E) \ 2 \pi \alpha \ d\alpha}
		{ F_a (1 - e^{-\tsca}) }.
\end{equation}
Multiplying Equation~\ref{eq:fh_fps} and Equation~\ref{eq:fcap} together, the observed flux ratio will be
\begin{equation}
	\label{eq:obs_fh_fps}
	\frac{\Fh^{cap}}{\Fps} = e^{\tsca} \int_{\alphmin}^{\alphmax} \frac{d\psi_h}{d\Omega} (\alpha, E)\ 2 \pi \alpha \ d\alpha.
\end{equation}
So in order to compute $\Fh^{cap}/\Fps$, one must compute a scattering halo model to account for missing flux.  A field of view on the order of $10-20'$ is necessary to capture the vast majority of halo light, thereby obtaining a model independent measurement for $\tsca$.

We split ObsId 6601 into energy separated images using 0.5~keV wide bins going from 1 to 6~keV.  $\Fh^{cap}$ was calculated by summing the flux in each annulus, after applying the same PSF and background subtraction methods described in Section~\ref{sec:Observation}, 
\begin{equation}
	\label{eq:FluxSum}
	\Fh^{cap} (\Delta E) = \sum_{\alphmin \leq r \leq \alphmax} 
		\frac{\mathcal{SB}(r, \Delta E) - \mathcal{SB}_{\rm psf}(r,\Delta E)}
		{\mathcal{A}(r, \Delta E)}
\end{equation}
but only including annuli where the residual surface brightness was positive.  
The $\Fps$ value was obtained by integrating the HETG fit described in Section~\ref{sec:SpectralFit}.

\begin{figure}
\begin{center}
	\includegraphics[scale=0.49, trim=10 0 0 0]{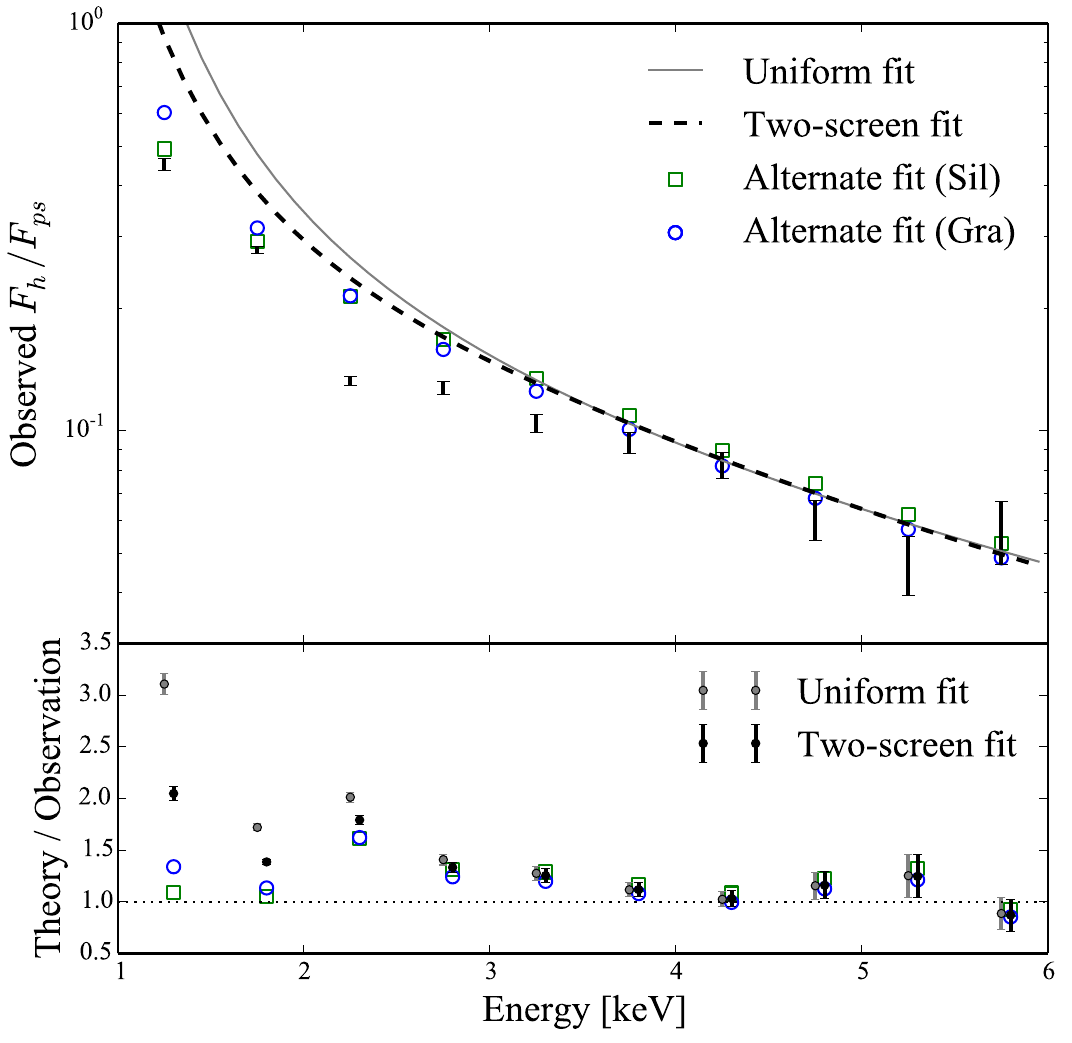}
	\caption{The ratio of halo flux ($F_h$, observed within $\approx 6'' - 100''$) to point source flux ($F_{ps}$) of Cyg X-3.  The solid grey and dashed black lines show theoretical expectations for the lowest $\bar{\chi}^2$ fit for uniform and two-screen dust distributions, respectively.  The bottom portion of the figure plots the ratio of theory to observation, showing that theoretical cross section over-predicts the amount of soft-energy scattering by a factor of 2-3.  The open circles and squares show the predicted ratios from Mie scattering calculations of graphite and silicate grains using an alternate fit to the 4-6~keV halo that incorporates large dust grains $\sim 1.5~\mum$ in the vicinity of the Cyg OB2 association (\S\ref{sec:AlternativeLargeGrains}).  
	}
	\label{fig:HaloRatio}
\end{center} 
\end{figure}

The grey and black dashed curves in Figure~\ref{fig:HaloRatio} show the observed flux ratios in comparison to that predicted from the 4-6~keV models fit in Section~\ref{sec:HaloFit}.  
For energy bins below 2.5~keV, the observed scattering halo flux is about a factor of two to three lower than that predicted from the scattering models used thus far.  We explore possible explanations for this behavior below, 
% --- Cut out sentences below, if I decide not to show alternative fit in Fig 7
including an alternative fit to the 4-6~keV scattering halo that utilizes large dust grains.  This alternative is plotted with open circles and squares in Figure~\ref{fig:HaloRatio} and will be explained below in Section~\ref{sec:AlternativeLargeGrains}.

%-------------------------
%-------------------------

%------------------
\subsection{Multiple Scattering}
\label{sec:MultipleScattering}

As shown in Sections~\ref{sec:SpectralFit} and \ref{sec:HaloFit}, the sight line to Cyg X-3 is optically thick to dust scattering for much of the energy range of interest.  
However, the calculations performed thus far only include the halo image from photons that scatter once through the intervening ISM.  Higher order scattering terms would alter the halo surface brightness profile in two ways: (i) increasing the intensity (Equation~\ref{eq:fh_fps}), and (ii) creating a more extended scattering image.  For more details, we refer the reader to \citet{ML1991}.

How do higher order scattering effects alter the predicted $\Fh^{cap}/\Fps$ curves?  Taylor expanding Equation~\ref{eq:fh_fps} gives an estimate for the relative flux contribution from each higher order halo: $F_{h1} \propto \tsca$, $F_{h2} \propto \tsca^2/2$, and so on.  Inclusion of higher order scattering terms would thereby {\sl increase} model $\Fh/\Fps$ by a factor $\sim \tsca/2$, which does not resolve the behavior of the data points in Figure~\ref{fig:HaloRatio}.

In addition, \citet{Smith2006} show that the second order scattering halos for MRN dust and $N_{\rm H} = 4 \times 10^{22}$ cm$^{-2}$ increases the intensity of the inner halo ($\alpha < 100''$) only by about 5\% and has little effect on the profile shape in that region.  
This means that our $F_h^{cap}$ value is likely accurate to within $5-10\%$.  
Note also that, in Equation~\ref{eq:fcap}, the denominator holds true for the optically thick case; higher order scattering terms will only affect the numerator in the equation.  From these observations we conclude that our predicted $\Fh^{cap}/\Fps$ curves would not change significantly with the inclusion of higher order scattering effects.

%-------------------------
\subsection{Scattering Contribution from Large Grains}
\label{sec:MieScattering}

%-------------------------
\begin{figure}
\begin{center}
	\includegraphics[scale=0.62, trim=0 0 0 0]{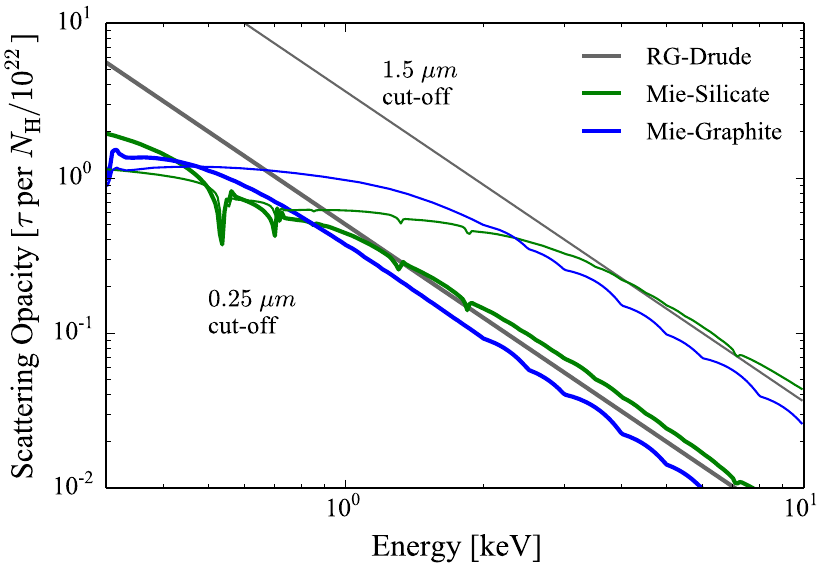}
	\caption{ Comparison between the RG-Drude (grey) and Mie scattering (blue and green) optical depth per $\NH = 10^{22}~\col$ for an MRN distribution of dust grains with $\amax = 0.25~\mum$, typical for Milky Way diffuse ISM (thick lines).  When $a_{\mum} \gsim E_{\rm keV}$, the RG-Drude approximation is no longer valid, and the Mie scattering solution shows that the scattering cross-section is significantly reduced at softer energies.  An experimental dust grain size distribution with the same power-law slope but $\amax = 1.5~\mum$ shows how increasing the grain size can cause the dust scattering optical depth to be reduced significantly for $E < 3$~keV (thin lines). }
	\label{fig:MieScattering}
\end{center} 
\end{figure}
%-------------------------

The downturn in observed $\Fh^{cap}/\Fh$ data points at low energies may be indicative of a population of large dust grains.  This is because the RG-Drude approximation breaks down when the grains become too large in comparison to the photon energy, violating the general rule-of-thumb $a_{\mum} \lsim E_{\rm keV}$ \citep{SD1998}.  
Figure~\ref{fig:MieScattering} shows how the Mie scattering solution \citep{BHbook} for two commonly hypothesized grain materials, silicate and graphite \citep{Draine2003b}, departs from the RG-Drude approximation at low photon energies.  Using a power law slope of -3.5, the Mie scattering solutions produce much lower cross-sections for $E \lsim 1$~keV when the grain size cut-off is $\amax = 0.25~\mum$, or for $E \lsim 3$~keV when $\amax = 1.5~\mum$.

There is reason to believe that grains larger than a few tenths of a micron exist along the Cyg X-3 sight line.  
The cold and dense environments of molecular clouds, such as the foreground Cygnus X molecular region, are likely to be a location of grain growth \citep{Draine2003}.  Evidence for micron-scale grains in molecular cloud cores has been observed through the infrared scattering phenomenon, or ``coreshine'' \citep[e.g.~][]{Steinacker2010,Andersen2013}.  While not likely associated with Cygnus X, the existence of a Bok globule imaged with X-ray scattering \citep{McCollough} also offers circumstantial evidence that large grains may obscure the sight line of Cyg X-3.  
In addition, the foreground young stellar association Cyg OB2 contains a number of Wolf-Rayet and candidate evolved stars \citep{Comeron2002}.  These types of stars experience significant mass loss in the form of dusty winds, causing large amounts of extinction, $A_V = 3 - 10$ \citep{Wright2015}.  The stellar winds of evolved stars are also likely sites for grain growth, and there is some evidence for micron-sized grains in evolved circumstellar environments \citep{deVries2015}.

Unfortunately, calculating the halo intensity with Mie scattering is computationally intensive, preventing a Bayesian analysis of the full range of energy resolved scattering halos at this time.  
A comparison of the Mie scattering halo intensity using the best fit parameters of Sections~\ref{sec:UniformFit} and \ref{sec:ScreenFit} are shown in Figures~\ref{fig:UniformTriangle} and \ref{fig:TwoScreenTriangle}.  As expected from Figure~\ref{fig:MieScattering}, the Mie scattering halos from dust distributions of pure silicate or pure graphite grains straddle the RG-Drude approximation.
In Section~\ref{sec:AlternativeLargeGrains} we use RG-Drude to explore a fit to the 4-6~keV halo that includes micron-sized grains, then extrapolate that fit to the lower energy band with Mie scattering.

%---------------------
\subsection{Dust optical properties}
\label{sec:Dielectrics}

Another possibility is that some population of dust grains in the foreground of Cyg X-3 have different optical properties than those used in this work \citep[graphite and astrosilicate from][]{Draine2003b}.  This could include composite grain types \citep{Zubko2004}, core-mantle grains \citep{LG1997}, or elongated grains \citep[e.g.][]{Min2003}.  Porous, ``fluffy'' grains will reduce the scattering cross-section because they have a lower material density ($\rho_3$ in Equation~\ref{eq:red_Sigma}).  However, the cross-section should still roughly follow the $E^{-2}$ power law for instances where the RG-Drude approximation holds.  Fluffy grains are thereby expected to  change the normalization but not the qualitative behavior of the curves in Figure~\ref{fig:MieScattering}.

Regardless, the data in Figure~\ref{fig:HaloRatio} show that the soft X-ray extinction properties of interstellar dust differ significantly from the often used RG-Drude approximation.  A future study incorporating a wider field-of-view, encompassing the entire scattering halo of Cyg X-3, would provide a direct measurement of the energy dependence of the dust scattering cross-section.  

%---------------------
\subsection{Time variation and pileup in Cyg X-3}
\label{sec:variability}

Cyg X-3 is a binary exhibiting deep absorption minima on a $4.8$ hour period \citep{CygX3period}, making it a good object for studying scattering in the time domain.  Since the scattered light takes a longer path, there is a delay between non-scattered and scattered light.  \citet[][]{Predehl2000} used the first {\sl Chandra} observations of Cyg X-3 to determine a geometric distance of $9^{+4}_{-2}$ kpc.  
ObsId 6601 is 49.6 ks long and covers 2.9 cycles of the Cyg X-3 period, as opposed to the 12.3~ks observation used by \citet{Predehl2000}, which covers only 0.7 cycles.  Given that ObsId 6601 contains nearly an integer number of cycles, we used the time-integrated image under the assumption that the average brightness of the point source describes the average brightness of the scattering halo.

To check that this assumption is valid, we divided ObsId 6601 into time intervals when Cyg X-3 was brightest (1-$\sigma$ above the mean count rate, for a total of 10.76~ks) and dimmest (1-$\sigma$ below the mean count rate, for a total of 11.35~ks).  We checked each time interval for pileup.  Based on the count rate in the 4-5~keV region, we expect that roughly 7\% of the HEG events are lost due to pileup during the brightest phase, and roughly 2\% during the dimmest phase.  Such low pileup fractions indicate that the behavior of the instrument will be nearly linear.  

The shape of the Cyg X-3 spectrum also varies with brightness.  There is more absorption from the stellar companion in the dim phases of the light curve in comparison to the bright phases.  Despite these differences, we found that averaging the spectrum from bright and dim phases yields a curve that is identical to the time-integrated spectrum, to within a few percent.  

Time variations in the scattering halo will only be visible for angles where the time lag is less than the period of the source.  In the case of Cyg X-3, this is true for observation angles $\alpha \leq 40''$.  When we extracted the radial profile from bright and dim time intervals, we indeed found variations in the soft-energy halo out to $\alpha \sim 20''$.  
However, given that (i) the spectral variations of Cyg X-3 scale linearly, and average out to the time integrated spectrum; and (ii) time variations in the scattering halo cannot exceed the period of Cyg X-3; and (iii) the duration of ObsId 6601 covers three whole cycles of the binary period, we conclude that the average time integrated flux computed in Section~\ref{sec:SpectralFit} should describe the average time integrated scattering halo in Section~\ref{sec:HaloFit}.

%---------------------
\subsection{Possible Sources of Systematic Error}
\label{sec:SystematicError}

Only two of the data points, one at 4.75~keV and the other at 5.25~keV, are systematically low because they contain some negative residual surface brightness values, causing a few annular bins to be ignored from Equation~\ref{eq:FluxSum}.

The abrupt decrease in scattering halo flux may be partly due to instrumental effects.  The Al-K and Ir-M absorption edges cause swift changes in the telescope effective area around 1.55 and 2.1 keV, respectively.  We did our best to correct for this by using weighted exposure maps of the source spectrum to build the PSF templates (Section~\ref{sec:Observation}).  Si-K and Au-M absorption edges (1.8 and 2~keV) within the CCD and HETG instruments themselves can also remove or alter, through fluorescence, the observed energy of incoming photons in this range.  This might contribute to the fact that the 1.75~keV and 2.25~keV energy bins exhibit the most dramatic discrepancies between model and data.

%----------------------------------
\section{Exploration of alternative fits}
\label{sec:Alternative}
%----------------------------------

\begin{table}
\centering
\caption{Alternative fits to the 4-6~keV dust scattering halo}
\label{tab:Alternative}
\begin{tabular}{l c c c c c}

	\hline	
	\multicolumn{6}{l}{ {\bf Two screen fit with all parameters free}} \\
	($\bar{\chi}^2 = 0.7$)		& $x$	& $\NH$~(a)	& $\amax$~(b)	& $p$	& $\tsca~E^2$~(c)
	\vspace{0.03in} \\
	\hline

	Screen 1: &
	$0.73$	& $1.0$				& $0.25$			& $1.6$ 	& $1.4$\\
	Screen 2: &
	$0.18$	& $0.6$				& $0.25$			& $3.4$	& $0.4$\\
%	Total: & 
%			& 1.6					& 				&		& 1.8 \\
	\hline	
	\multicolumn{6}{l}{{\bf Two screen fit with large grains in Cyg OB2}} \\
	($\bar{\chi}^2 = 0.6$)	 & $x$	& $\NH$~(a)	& $\amax$~(b)	& $p$	& $\tsca~E^2$~(c)
	\vspace{0.03in} \\
	\hline

	Screen 1: &
	0.45		& 3.0					& 0.18*			& 3.5*	 & 1.6 \\
	Screen 2: & 
	0.78		& 0.08				& 1.12			& 3.0		& 0.3 \\ 		
%	Total &
%			& 3.08				& 				& 		& 1.9 \\
	\hline
	\multicolumn{2}{l}{(a) $10^{22}$~cm$^{-2}$} & 
	\multicolumn{1}{l}{(b) $\mum$} & 
	\multicolumn{3}{l}{(c) keV$^{2}$} \\
	\multicolumn{6}{l}{* Fixed model parameter} \\
	
\end{tabular}
\end{table}

We explored a number of alternative fits to the 4-6~keV scattering halo, in attempts to explain the energy resolved scattering halos evaluated above.  None of them were able to completely fit the scattering halo behavior in the 1-6~keV range.  Nonetheless, we document them here for completeness.  Table~\ref{tab:Alternative} gives a summary of the best fit parameters for alternative two-screen dust scattering halo models.

%-------------------------------------
\subsection{Two screens with different grain size distributions}

We performed a least-squares fit with no priors to the 4-6~keV scattering halo, allowing all parameters to vary so that each dust screen was allowed to have its own grain size distribution.  This requires 60\% of the dust to be in a foreground screen with a shallow grain size distribution ($p=1.6$), so that the majority of the mass is in $0.25~\mum$ grains.  The other 40\% of the dust follows an MRN distribution and is contained in a screen close to Cyg X-3.

We checked to see if the noticeably shallow grain size distribution for Screen~1 turned away from the RG-Drude approximation at energies larger than 1~keV.  The resulting $\tsca$ curve did not differ significantly from the behavior seen in Figure~\ref{fig:MieScattering}, showing that the total X-ray scattering cross-section is more sensitive to $\amax$ than $p$.  As with the fits presented in Section~\ref{sec:HaloFit}, this model over-predicts the amount of scattering at soft energies.

%-------------------------------------
\subsection{Large grains in the vicinity of Cyg OB2}
\label{sec:AlternativeLargeGrains}

Inspired by the Mie scattering cross-sections in Figure~\ref{fig:MieScattering}, we explored an alternate fit to the 4-6~keV halo that replaces the far screen of dust grains ($x_2 \approx 0.1$) with a hypothesized screen of large grains coextensive with the Cyg OB2 association ($x_2 \approx 0.8$).  This nearby scatterer needs to contain grains $\sim 1-2~\mum$ in radius to form the inner portion of the scattering halo, which is mainly confined to a $10''$ region.  Properties of the dust grain size distribution in the first screen, corresponding to scattering from the Perseus arm, were held fixed using values from Section~\ref{sec:ScreenFit}.  
In this model the screen associated with Cyg OB2 contains only a few percent of the total dust mass and accounts for about 20\% of all scattering for $E > 4$~keV (where the RG-Drude approximation holds).  The posterior distribution median and 1-$\sigma$ confidence interval for the grain size cut-off is $\amax = 1.5^{+ 0.5}_{-0.4}~\mum$.

\begin{figure}
\begin{center}
	\includegraphics[scale=0.6, trim=0 0 0 0]{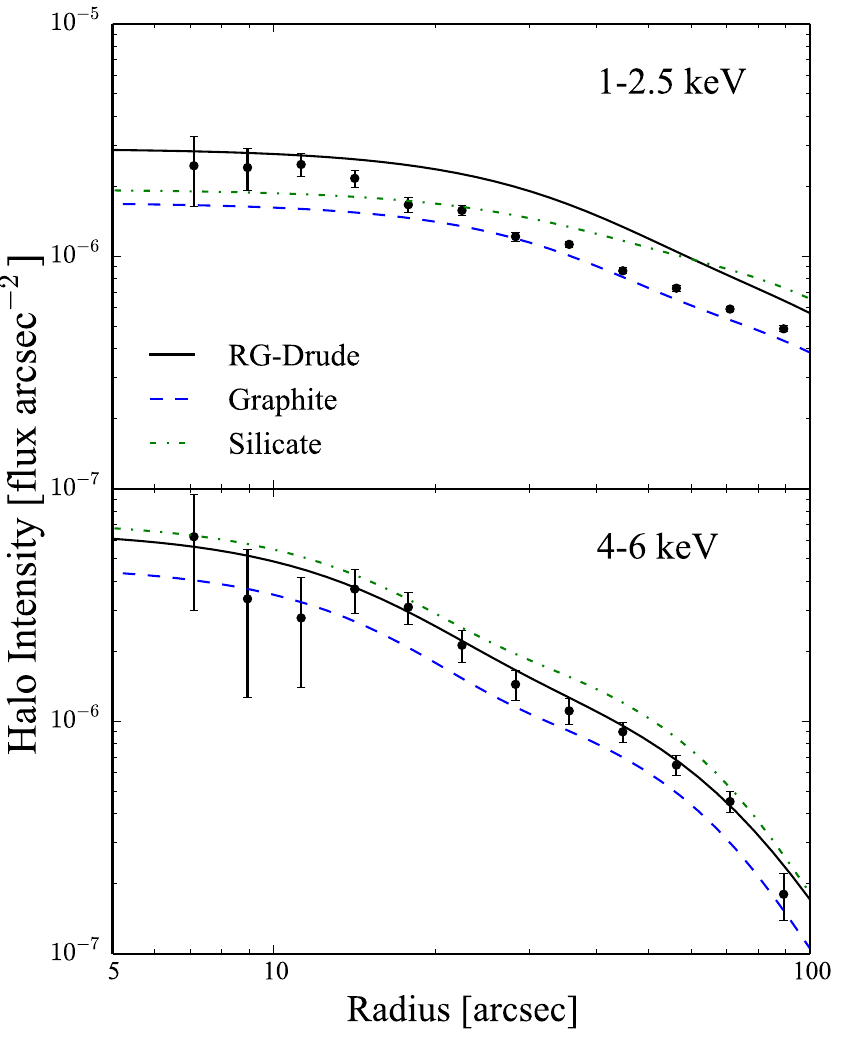}
	\caption{ X-ray scattering halo intensities from a fit that includes large micron-sized dust grains in the vicinity of the Cyg OB2 association.  The 1-2.5~keV curves ({\sl top}) were extrapolated from the RG-Drude fit to the 4-6~keV scattering halo intensity ({\sl bottom}).  In situations where the RG-Drude approximation is appropriate, the Mie scattering solution for a population of purely silicate (green dashdot) or purely graphite (blue dash) grains straddles the RG-Drude solution (black).  In the lower energy band where RG-Drude does not hold, the Mie scattering solution produces dimmer scattering halos overall.}
	\label{fig:LargeDustScattering}
\end{center} 
\end{figure}

Figure~\ref{fig:LargeDustScattering} shows the scattering halo intensity for the best fit obtained from the posterior distribution (Table~\ref{tab:Alternative}) in the 1-2.5~keV and 4-6~keV bands.  The RG-Drude approximation, used to fit the 4-6~keV halo intensity, is shown for the 1-2.5~keV band as a comparison to the Mie scattering solutions.  By including a thin screen of large grains in the foreground, the inner portion of the soft-energy scattering halo drops out by a factor of about two.  There is still a noticeable systematic offset in the the outer portion of the scattering halo profile, which is dominated by the more distant (Perseus arm) screen that contains MRN-type dust.  The fact that this dust model does not reconcile the energy-resolved scattering halos entirely emphasizes the need for testing the energy dependence of the dust scattering cross-section over a broader range of X-ray energies.

%-------------------------
%-------------------------

To incorporate this fit into Figure~\ref{fig:HaloRatio}, we computed the scattering halos at lower energies using the Mie scattering solution for Screen 2 (Cyg OB2) and the RG-Drude approximation for Screen 1 (Perseus arm).  Calculating the Mie scattering halos is computationally intensive, so we computed them using only the bin centers for each 0.5~keV wide energy band.  Figure~\ref{fig:HaloRatio} illustrates that the inclusion of micron-sized grains brings the predicted halo flux into closer alignment with the observed values.  
It is also interesting to note that the median $\amax$ value for this scenario agrees with the models of \citet{Wang2014}, which show that $1.5~\mum$ sized dust grains can reproduce the mid-IR extinction curve of the Milky Way.
However, there is still a huge discrepancy in the 2-2.5~keV bin that remains a mystery.  The question of whether this is an instrumental effect or a true ISM absorption feature can be addressed by a study that compares dust scattering halos from multiple instruments.

%----------------------------------
\subsection{Implied Extinction}
\label{sec:Extinction}

The incorporation of large dust grains for modeling the X-ray scattering halo of Cyg X-3 begs the question: what will the extinction curve look like?  ISM extinction curves can vary across different regions of the Milky Way, but they are generally characterized by the magnitude of extinction relative to the color change: $R_V \equiv A_V / E(B-V)$ \citep{Cardelli1989}.   Extinction curves with a high $R_V$ can be produced by extending the dust grain population to larger sizes \citep[e.g.][]{WD2001}.  However, ZDA model distributions with large dust grains composed of amorphous carbon or composite materials can also reproduce the average Milky Way extinction curve of $R_V = 3.1$.

%-------------------------
\begin{figure*}
\begin{center}
	\includegraphics[scale=0.48, trim=0 0 0 0]{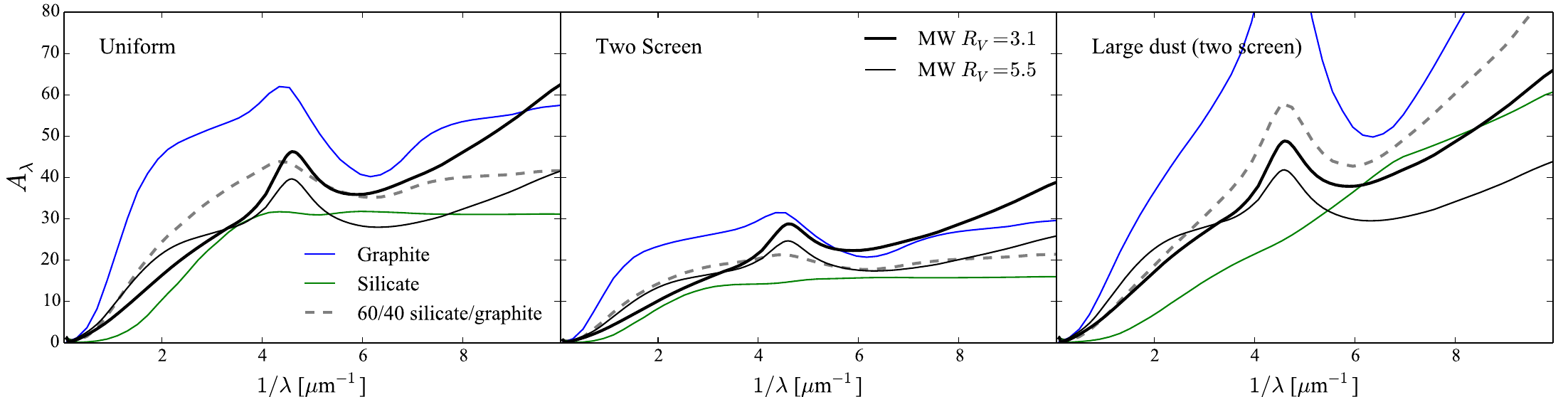}
	\caption{ The extinction curves utilizing the best fit grain size distributions are shown for the uniform ({\sl left}), two-screen ({\sl middle}), and large grain ({\sl right}) models for the X-ray scattering halo of Cyg X-3.  The true extinction values must be somewhere between that of pure graphite (blue) and silicate (green).    The synthetic extinction curves of \citet{WD2001} for $R_V=3.1$ and $R_V=5.5$ Milky Way dust are shown for comparison.  We use a 60\% silicate, 40\% graphite mix of grains (grey dashed line) to estimate the total $A_V$ towards Cyg X-3 for each model.
	}
	\label{fig:ExtinctionCurves}
\end{center} 
\end{figure*}
%-------------------------

Figure~\ref{fig:ExtinctionCurves} shows the extinction curves calculated from the best fit models in Sections~\ref{sec:UniformFit}, \ref{sec:ScreenFit}, and \ref{sec:AlternativeLargeGrains}.  We found that a 60/40 mix of silicate/graphite grains with an MRN size distribution of $p=3.5$ and $\amax = 0.3$ roughly matches the synthetic $R_V = 3.1$ extinction curve from \citet{WD2001}.  We use this mixture as a guess for the total extinction towards Cyg X-3, because at this time we cannot constrain the composition of our dust grains from the dust scattering halo alone.  The variation in total magnitude of extinction seen in Figure~\ref{fig:ExtinctionCurves} comes from the different dust masses (parameterized by $\NH$) implied by the fits.

In the cases of the best fit uniform and two-screen models, the extinction curve in the optical band more closely resembles the $R_V = 5.5$ curve.  This is due mainly to the fact that the best fit walkers in the distribution both have shallower power laws ($p<3$) than the median values, putting greater weight towards large grains.  The extinction curve calculated from the fit incorporating micron-sized grains more closely resembles the $R_V = 3.1$ curve because the majority of the dust mass is in the population of MRN-type grains held fixed with the Perseus arm.

The continuum fit to the X-ray spectrum of Cyg X-3, $\NH \sim 4 \times 10^{22}\col$, implies a total extinction of $A_V \sim 20$ using the relations of \citet{PS1995} and \citet{GO2009}.  The curves shown in Figure~\ref{fig:ExtinctionCurves} are roughly consistent with this value, but it depends strongly on the relative abundance of dust types.   The 60/40 mixture of silicate to graphite produces $A_V = 22$, 13, and 17 for the uniform, two-screen, and  large dust models, respectively.

%----------------------------------
\section{Comparison to other X-ray scattering halo studies}
\label{sec:Comparison}
%----------------------------------

\citet[][hereafter PS95]{PS1995} included Cyg X-3 in a study of X-ray scattering halos visible with ROSAT.  
The halo around Cyg X-3, being 40\% as bright as the central point source, had the largest optical depth in the study.
Since the ROSAT PSF is much broader, only the outer part of the halo ($\alpha > 100''$) was visible, leading them to be more sensitive to dust close to the observer.  Regardless, they got similar values for $\amax$ (0.20 $\mum$) and $p$ (3.8), which are consistent with our uniform fit.

The scattering halo of Cyg X-3 was also included in a broad study of high mass and low mass X-ray binaries \citep{Xiang2005}.  Using ObsID 425, they extracted the halo image from the 1-d projection of the halo data along the direction of the MEG arm \citep[see also][]{Yao2003}.  All of their binaries exhibited surface brightness profiles that are steep in the inner $5''$ (10 pixels) and less steep in the outer regions.  This led them to conclude that all of the objects in the study were embedded in molecular clouds, because all of the dust mass needed to be located very close to the X-ray sources.  In the case of Cyg X-3,  this required $N_{H} \sim 4 \times 10^{22}$ cm$^{-2}$ within $100$ pc of the HMXB.  Our study does not resolve the scattering halo within $5''$, but most of our two-screen fits do require a significant mass of dust within 1-2~kpc of Cyg X-3.  It is also possible that the particularly indirect method they used to resolve the scattering halo led to a systematic error in the inner regions of the halo.  We measure the scattering halo directly from the zeroth order image of HETG TE-mode data instead.

\citet{Ling2009} performed an in depth cross-correlation timing analysis of Cyg X-3 to determine the spatial distribution of dust.  They found that dust associated with Cyg OB2 could account for time lags in the $\alpha > 65''$ halo, but that only accounted for 7\% of the total dust along the line of sight.  The remainder of the data was well fit with uniformly distributed dust. We do not perform a uniform plus screen fit because the screen contribution would be similarly negligible.

Depending on the Cyg OB2 distance, \citet{Ling2009} measured 3.4, 7.2, and 9.3 kpc to Cyg X-3.
We put rather strict priors on the screen positions to correspond to the Perseus and outer spiral arms of the Milky Way, because the scattering halo from MRN-type dust in the nearby Cyg OB2 cluster would not be contribute enough to the $\alpha \leq 100''$ scattering halo to substantially affect our fits.  
For the sake of thought experiment, associating Screen~1 with the Cyg OB2 cluster at $1.4 \pm 0.08$~kpc \citep{Rygl2012} requires Cyg X-3 to be $2.5^{+0.6}_{-0.5}$~kpc away.  
If Screen~1 is associated with the Perseus arm $5.6 \pm 0.5$~kpc away \citep[Figure~1 of][]{Reid2014}, then our posterior distribution from Section~\ref{sec:ScreenFit} implies that Cyg X-3 is $9.8^{+2.4}_{-2.1}$~kpc away, which is consistent with \citet{Predehl2000}.
 
%----------------------------------
\section{Conclusion}
\label{sec:Conclusion}
%----------------------------------

In the Appendix we present an analytic solution for the dust scattering halo intensity from a power law distribution of grain sizes in the single scattering RG-Drude regime.  The solutions are in the form of erf and incomplete gamma functions, which are included in many common software packages such as Scientific Python and Mathematica.  This allows for a probabilistic approach to fitting the data, which required computing $\sim 10^5$ halos for each of our dust distribution models.  The Bayesian approach is powerful because it allows us to explore and constrain the degeneracies that exist between dust grain size, power law slope, and spatial distribution that produce scattering halo surface brightness profiles.  The Bayesian analysis also allowed us to incorporate and quantify prior knowledge about the relative position of two dusty foreground screens.

We take this approach to analyze one of the brightest dust scattering halos available in the {\sl Chandra} archive, that associated with Cyg X-3. We find that a uniform distribution of dust along the line of sight fits the scattering halo profile for the region $\alpha < 100''$.  This suggests an MRN-type grain size distribution with a slightly smaller grain size cut off than typically assumed, $\amax = 0.18 \pm 0.03~\mum$.  The scattering halo can also be fit with two infinitesimally thin dust screens placed in the foreground of Cyg X-3.  About 80\% of the dust would be located about half-way along the sight line and is most likely associated with the Perseus spiral arm of the Milky Way.  The remaining 20\% of the dust would have to be within 1~kpc of Cyg X-3, contributing most to the inner portion ($r<15''$ pixels) of the scattering halo.  The grain size distribution suggested by the two-screen fit has a similar cut off to the uniform model ($\amax = 0.17^{+0.06}_{-0.04}~\mum$).  Our results are consistent with other published conclusions regarding the distance to Cyg X-3 and the dust grain distribution along its sight line.

%We took this approach to fitting the 4-6~keV scattering halo of Cyg X-3, a HMXB with complex foreground structures.  
%%We have shown that the intensity of a dust scattering halo arising from a power law grain size distribution can be computed quickly, allowing for a Bayesian MCMC fit to the 4-6~keV scattering halo of Cyg X-3.  
%%Because many degeneracies exist between dust grain size, power law slope, and dust spatial distribution, a probabilistic approach to these fits is favorable.
%%
%Since the RG-Drude approximation is strongly sensitive to dust grain size, it generally probes the upper limit of more complicated grain size distributions such as WD01 and ZDA.  
%%
%Our approach attempts to derive the grain size upper limit from the data.  In the future, exponential decay around the grain size cut-off can be implemented in order to contribute to, not just test, more advanced grain size distributions.

%We have evaluated the Bayesian Information Criterion to find that a model implementing uniformly distributed dust along the line of sight (Section~\ref{sec:UniformFit}, BIC$=16.0$) performs similarly to a model of two dusty screens associated with the Perseus and outer Milky Way spiral arms (Section~\ref{sec:ScreenFit}, BIC$=18.4$).  

The fact that a power law distribution of dust can well describe the shape of the 4-6~keV scattering halo attests to the survival of the MRN model, which has a noteworthy ability to describe Milky Way dust {\sl on average}.  This is surprising when we consider the variety of interstellar environments expected along the Cyg X-3 sight line, which includes young stellar clusters and molecular material in addition to the diffuse ISM.  
However, when we look at the broad band energy resolved scattering halos from 1 to 6~keV, the power law distribution of dust grains fails due to the shortcomings of the RG-Drude approximation, the dust optical constants used, or both.  
The nature of the resolved soft energy $E \lsim 2.5$~keV scattering halos, which are much dimmer than expected when extrapolating from the 4-6~keV fits, might be explained by the reduced scattering efficiency of soft energy X-rays by micron-sized grains.  The molecular regions, foreground Bok globule, extinction properties, and evolved nature of many Cyg OB2 cluster members in the angular vicinity of Cyg X-3 %emphasize the unique nature of this particular sight line, which might host 
suggest that this sight line might host dust grains of larger size or different composition than those in the diffuse ISM.  However, we are unable to resolve this issue in the single scattering RG-Drude regime. 

%We hypothesize that large dust grains in the foreground of Cyg X-3 might be able to explain the relative dimness of the soft energy halos with respect to the 4-6~keV model.  An exploration of dust optical properties and incorporation of Mie scattering into the fitting process is reserved for a future work.

%In this paper, we showed how Bayesian analysis allowed us to incorporate and quantify prior knowledge about the relative position of the two dusty foreground screens.  
%
%While this model performs similarly to the uniform fit, we find that there is less discrepancy between the two screen fit and the 1-2.5 keV residuals for $r < 30$ pixels (Section~\ref{sec:BigGrains}).  Thus an optically thin body of dust within 1 kpc of the Cyg X-3 best describes the inner portion of the scattering halo.  

Finally, we would like to point out from the literature that all current interstellar grain models do not fit observed X-ray scattering halos with $\bar{\chi}^2 < 2-5$ \citep[e.g.][]{Smith2006, Valencic2008}, unless a significant number of free parameters are employed by incorporating multiple dust clumps in the ISM spatial distribution, as in this work and others \citep[e.g.][]{Valencic2009, Xiang2011}.  
Since the RG-Drude approximation is strongly sensitive to dust grain size, it generally probes the upper limit of more complicated grain size distributions such as WD01 and ZDA.  
Our approach attempts to derive the grain size upper limit from the data.  In the future, exponential decay around the grain size cut-off, dual power laws, or other parameterized distributions should be implemented in order to contribute to, not just test, our current understanding of grain size distributions and growth.

This work also presents resolved X-ray scattering halos over a wider energy band than typically covered in the literature.  Fits utilizing the RG-Drude approximation and single scattering regime differ significantly when applied to $E \lsim 2.5$~keV as opposed to those applied to higher energy bins, where the approximations are more appropriate.  We encourage future works treating dust scattering along optically thick $N_{\rm H} \gsim 10^{22}~\col$ sight lines to examine $E \gsim 2.5$~keV scattering.
We also suggest that future researchers use the Mie scattering cross-section for $E \lsim 2.5$~keV scattering halos from optically thick sight lines.

%---------------------------------
\section*{Acknowledgements}

The python code used to model the scattering halos and extinction curves is available online at http://github.com/eblur/dust \citep{dustrepo}.  
This research has made use of the SIMBAD database, operated at CDS, Strasbourg, France.
This work was supported in part by NASA Headquarters under the NASA Earth and Space Science Fellowship Program, grant NNX11AO09H.

We thank Michael McCollough, R.~K.~Smith, L.~Valencic, D.~Foreman-Mackey, Herman L.~Marshall, John Davis, John Houck, and the entire CXC staff at MIT for useful discussion and calibration advice.  

\bibliography{references,CygX3}

\appendix
\onecolumn
\section{Solutions for dust scattering from a power law distribution of grain sizes}

The integral for the scattering halo intensity can be evaluated analytically under a few simple conditions.  First, we assume that the grain size distribution is a power law function of grain size:
\begin{equation}
\label{eq:dust_powerlaw}
	N_d \propto a^{-p}
\end{equation}

Second, we use the Rayleigh-Gans differential cross section as described in the text (Equation~\ref{eq:red_dSigma}):
\begin{equation}
\label{eq:appendix_dSigma}
	\frac{d\sigma}{d\Omega} \propto a^6\ 
	\exp\left(  \frac{-\alpha^2 a^2}{2 \charsig_0^2 x^2} \right)
\end{equation}
where $\charsig_0$ is the characteristic scattering angle for 1 $\mum$ size grains, such that $\charsig_0 = 1.04' \ E_{\rm keV}^{-1}$.

Finally, we assume that the medium along the line of sight is optically thin to dust scattering.  The single-scattering halo intensity, integrated over solid angle, is
\begin{equation}
\label{eq:thin_equation}
	\int I_h d\Omega = F_a \tau_{\rm sca}
\end{equation}
For more information on second or third order scattering see \citet{ML1991}.

We can integrate Equation~\ref{eq:halo_intensity} over solid angle to solve for one of our normalization factors.  We will use $A$ as a normalization constant that combines the dust grain size distribution and differential cross-section proportionalities described above.
\begin{equation}
	F_a \tau_{\rm sca} = \ F_a A \ 
	\int a^{6-p} \ \int x^{-2} \xi(x) \
	\int_0^\infty \exp\left( -\frac{ \alpha^2 }{ 2 \charsig_0^2 } \frac{ a^2 }{ x^2 } \right) \ 
	2\pi \alpha \ d\alpha \ dx \ da
\end{equation}
Integrating over $\alpha$ first will contribute an $(x/a)^2$ term, and we will drop the $\xi(x)$ term for now.  This yields
\begin{equation}
\label{eq:A}
	A = \frac{ \tau_{\rm sca} }{ 2 \pi \charsig_0^2 \ G_p(a,p) }
\end{equation}
where $G_p$ is a constant:
\begin{equation}
\label{eq:G_p}
	G_p(a,p) \equiv \int_{\amin}^{\amax} a^{4-p} da
\end{equation}

Under the above simplifying notation, Equation~\ref{eq:halo_intensity} becomes
\begin{equation}
\label{eq:halo_intensity_appendix}
	I_h(\alpha) = 
	\frac{ F_a \ \tau_{\rm sca} }{ 2 \pi \charsig_0^2 \ G_p(a,p) } \  
	\int a^{6-p} \ \int x^{-2} \xi(x) \
	\exp\left( -\frac{ \alpha^2 }{ 2 \charsig_0^2 } \frac{ a^2 }{ x^2 } \right) \ 
	dx \ da
\end{equation}

\subsection{Screen case}

In the case of an infinitesimally thin screen at position $x_s$, Equation~\ref{eq:halo_intensity_appendix} becomes
\begin{equation}
	I_h(\alpha) = \frac{ F_a }{ x_s^2 } \ 
		\frac{ \tau_{\rm sca} }{ 2 \pi \charsig_0^2 \ G_p(a,p) }
		\int a^{6-p} \exp \left( \frac{ -\alpha^2 a^2 }{ 2 \charsig_0^2 x_s^2 } \right)
		\ da
\end{equation}
which produces the solution
\begin{equation}
	 I_h(\alpha) = \frac{F_a}{x_s^2} \
	 \frac{ \tau_{\rm sca} }{ 2 \pi \charsig_0^2 } \
	 \frac{ G_s(a,p,\alpha,x_s) }{ G_p(a,p) }
\end{equation}
where
\begin{equation}
\label{eq:G_s}
	G_s(a,p,\alpha,x_s) \equiv -\frac{1}{2} \left[ \left(\frac{\alpha^2}{2 \overset{\sim}{\sigma_0}^2 x_s^2} \right) ^{\frac{p-7}{2}} \Gamma\left( \frac{7-p}{2}, \frac{\alpha^2 a^2}{2 \overset{\sim}{\sigma_0}^2 x_s^2} \right) \right] _{\amin}^{\amax} 
\end{equation}

\subsection{Uniform case}

In the case that the dust grains are uniformly distributed along the line of sight, Equation~\ref{eq:halo_intensity_appendix} becomes
\begin{equation}
	I_h(\alpha) = 
	\frac{ F_a \ \tau_{\rm sca} }{ 2 \pi \charsig_0^2 \ G_p(a,p) } \  
	\int a^{6-p} \ \int x^{-2} \
	\exp\left( -\frac{ \alpha^2 a^2 }{ 2 \charsig_0^2 x^2 } \right) \ 
	dx \ da
\end{equation}
The $x$ term of the integral evaluates to
\begin{equation}
	\sqrt{\frac{\pi}{2}} \ \frac{\charsig_0}{\alpha a}
	\left[ 1 - {\rm erf} \left( \frac{ \alpha a }{ \charsig_0 \sqrt{2} } \right) \right]
\end{equation}
Plugging this in, we get
\begin{equation}
	I_h(\alpha) = 
	\frac{ F_a \ \tau_{\rm sca} }{ \alpha \charsig_0 \ \sqrt{8 \pi} \ G_p(a,p) } \  
	\int a^{5-p} \
	\left[ 1 - {\rm erf} \left( \frac{ \alpha a }{ \charsig_0 \sqrt{2} } \right) \right]
	\ da
\end{equation}

The solution is
\begin{equation}
	I_h = \frac{F_a}{\alpha \charsig_0}\ 
	\frac{\tau_{\rm sca}}{ \sqrt{8\pi} }\ 
	\frac{G_u (a,p,\alpha) }{G_p (a,p)}
\end{equation}
where
\begin{equation}
\label{eq:G_u}
	G_u(a,p,\alpha) \equiv \frac{1}{6-p} \left[ a^{6-p} 
	\left( 1 - {\rm erf} \left( \frac{\alpha a}{\overset{\sim}{\sigma_0} \sqrt{2}} \right) \right)
	- \frac{1}{\sqrt{\pi}} \left( \frac{\alpha}{\overset{\sim}{\sigma_0} \sqrt{2}} \right)^{p-6} 
	\Gamma\left( \frac{7-p}{2}, \frac{\alpha^2 a^2}{2 \overset{\sim}{\sigma_0}^2} \right)
	\right]_{\amin}^{\amax}
\end{equation}

\end{document}